\documentclass[12pt]{JHEP3}

\usepackage{epsfig}
\epsfclipon
\usepackage{multicol}
\usepackage{epsfig,bm}
\usepackage{amssymb,amsmath}

\newcommand{\roughly}[1]{\mathrel{\raise.3ex\hbox{$#1$\kern-0.85em
\lower1ex\hbox{$\sim$}}}}

\def\be{\begin{equation}}
\def\ee{\end{equation}}
\def\bea{\begin{eqnarray}}
\def\eea{\end{eqnarray}}
\def\hf{\frac12}
\def\pref#1{(\ref{#1})}

\def\cQ{{\cal Q}}
\def\cB{{\cal B}}

\newcommand{\bmat}{\left(\begin{array}}
\newcommand{\emat}{\end{array}\right)}

\def\yzero{\smash{\hbox{$y\kern-4pt\raise1pt\hbox{${}^\circ$}$}}}

\def\beq{\begin{equation}}
\def\eeq{\end{equation}}
\def\beqa{\begin{eqnarray}}
\def\eeqa{\end{eqnarray}}
\def\-{\hphantom{-}}

\def\s2{\frac{1}{2}}

\def\beq{\begin{equation}}
\def\eeq{\end{equation}}
\def\beqa{\begin{eqnarray}}
\def\eeqa{\end{eqnarray}}

\def\IF{\relax{\rm I\kern-.18em F}}
\def\II{\relax{\rm I\kern-.18em I}}
\def\IP{\relax{\rm I\kern-.18em P}}
\def\IC{\relax{\rm I\kern-.48em C}}
\def\IR{\relax{\rm I\kern-.18em R}}
\def\IK{\relax{\rm I\kern-.20em K}}
\def\IM{\relax{\rm I\kern-.25em M}}
\def\KKLMMT{{\IK L\IM T}}
\def\KKLT{{\IK LT}}

\def\cp{{\cal P}}

\def\Dsl{\,\raise.15ex\hbox{/}\mkern-13.5mu D} %this one can be subscripted
\def\IZ{{\Bbb Z}}
\def \one{\relax{\rm 1\kern-.26em I}}

\def\cQ{{\cal Q}}

 \def\cp#1{\relax\ifmmode {\IP\kern-2pt{}_{#1}}\else $\IP\kern-2pt{}_{#1}$\=fi}
\def\ReT{\sigma} %%CB2: this keeps Re T = sigma/2
\def\ReB{b}  %%CB2: this is to make Re B = b/2 (this was already used in chap 2)
\def\ImB{\beta}  %%CB2: this is to make Im B = beta
\def\ImT{\tau}  %%CB2: this is to make Im T = tau
\def\ArgPhi{\theta} %%CB2: this is to make arg phi = theta
\def\ModPhi{\psi} %%CB2: this is to make |phi| = \psi

\title{Inflation in Realistic D-Brane Models}

\author{C.P.\ Burgess,$^1$ J.M. Cline,$^1$ H. Stoica$^1$ and F. Quevedo$^2$ \\
        $^1$Physics Department, McGill University \\
        3600 University Street, Montr{\'e}al \\
        Qu{\'e}bec, Canada, H3A 2T8. \\

$^ 2$DAMTP, Centre for Mathematical Sciences\\
               University of Cambridge,\\
               Cambridge CB3 0WA UK.}

\date{}
\abstract{We find successful models of D-brane/anti-brane
inflation within a string context. We work within the GKP-\KKLT\
  class of
type IIB string vacua for which many moduli are stabilized through
fluxes, as recently modified to include `realistic' orbifold
sectors containing standard-model type particles. We allow all
moduli to roll when searching for inflationary solutions and find
that inflation is not generic inasmuch as special choices must be
made for the parameters describing the vacuum. But given these
choices inflation can occur for a reasonably wide range of initial
conditions for the brane and antibrane. We find that D-terms
associated with the orbifold blowing-up modes play an important
role in the inflationary dynamics. Since the models contain a
standard-model-like sector after inflation, they open up the
possibility of addressing reheating issues. We calculate
predictions for the CMB temperature fluctuations and find that
these can be consistent with observations, but are generically not
deep within the scale-invariant regime and so can allow
appreciable values for $dn_s/d\ln k$ as well as predicting a
potentially observable gravity-wave signal. It is also possible to
generate some admixture of isocurvature fluctuations.}

\preprint{DAMTP-2003-141, McGill-04/04}


\begin{document}

%\newpage

%===================================================================================

\section{Introduction}
The possibility of having cosmological inflation arise due to the
relative motion of D-branes and their anti-branes is very
attractive \cite{bmnqrz,dss,morebi}.\footnote{See also
\cite{stephon} for an early brane-antibrane proposal which does
not rely on the relative inter-brane motion as the inflaton and
\cite{angles} for extending the brane/antibrane system to branes
at small angles.} It provides an explicit and geometrical
interpretation of the inflaton field as the separation of the
D-branes \cite{dt}, with slow roll potentially achieved through a
calculably weak effective attraction. It also includes a naturally
graceful exit from inflation due to the necessary appearance of an
open string tachyon at a critical separation \cite{bmnqrz},
providing a stringy realization of the hybrid inflation
\cite{hybrid} scenario. This potentially permits further
connections between cosmology and string theory through the
properties of the tachyon field which have been recently
discovered \cite{sen}.

Its great promise as an inflationary mechanism has sharpened the
search for an explicit realization of this scenario within a
string compactification. This search has proven to be difficult,
for several reasons. First, as originally pointed out in
\cite{bmnqrz}, slow roll generically does not occur for brane
motion in compact spaces because the branes typically cannot get
far enough apart to let their interactions become sufficiently
weak. Compact spaces also raise another difficulty, since the
projecting out of bulk-field zero modes also makes slow rolls
difficult to achieve \cite{kklmmt}. (See \cite{jellium} for a
recent discussion of ways to avoid this last difficulty.)

A second serious obstacle has been the strong technical assumption
(made in all of the original proposals) that all string moduli but
the putative inflaton be fixed by some unknown string effect.
Recent progress circumventing this difficulty has come with the
realization that string moduli can be explicitly fixed if the
extra dimensions are appropriately warped due to the presence of
fluxes \cite{kklt}. However even in this case inter-brane
inflation has been difficult to obtain \cite{kklmmt} due to a
variant of the standard $\eta$-problem of supergravity
inflationary models \cite{etaprob}. It is nonetheless expected
that inflation can occur in these vacua, although possibly at the
expense of fine tunings in the brane initial conditions to roughly
a part in 100.

Model-building suggests two features to seek in any inflationary
candidate within string theory. First, resolution of the $\eta$
problem suggests looking for a $D$-term inflationary mechanism
\cite{dterm, KS}, with the inflationary energy density being driven by
a supersymmetric $D$-term which is independent of (or weakly
dependent on) the putative inflaton. Second, successful
post-inflationary reheating requires the model be realistic in the
sense that it is possible to identify where Standard Model degrees
of freedom reside once inflation ends. The challenge is to find
real string vacua with these features, and for which as many
moduli are fixed as possible. This is the motivation for the
present paper, and we find that the inclusion of Standard Model
sectors automatically introduces $D$-term potentials.

We base our inflationary scenario on a recent extension to
realistic models \cite{cmqu} of the \KKLT\ mechanism \cite{kklt}
for moduli stabilization. The extension we use requires adding
extra branes where the (chiral) standard model particles can sit.
These models are particularly attractive for our purposes because
they incorporate many desirable features for phenomenology.
Besides including the spectrum of the standard model with three
chiral families of quarks and leptons, they also include a
mechanism for fixing the moduli and for generating a hierarchy
through warping, {\it \`a la } Randall and Sundrum
\cite{rs}-\cite{gkp}. The models considered necessarily require
more than one K\"ahler modulus to be present and therefore the
effective potential depends on more than the few fields of the
\KKLT\ scenario. We identify here the essential low-energy
features of this scenario in order to explore their prospects for
obtaining inflation.

We organize our presentation as follows. In order to set the stage
for our own work, in the next section we briefly summarize recent
developments, including both the string vacua which arise in the
\KKLT\ \cite{kklt} and \KKLMMT\ \cite{kklmmt} cosmological models
of recent interest and the construction \cite{cmqu} which allows
realistic string vacua to be embedded into these constructions.
Section 3 follows this with a description of the effective 4D
theory which captures the main features of the low-energy dynamics
of the moduli of these string vacua. In section 4 we follow
brane-antibrane motion with this low-energy moduli space in search
of inflationary slow rolls. In certain circumstances we are able
to identify sufficiently slow rolls and this section describes the
required circumstances in detail. In Section 5 we close with some
concluding remarks, including some words concerning reheating and
whether string theory may prefer to produce a comparatively short
period of inflation at the epoch of horizon exit for the largest
scales currently observed in the fluctuations of the cosmic
microwave background.

\section{Fluxes, Warping and  Moduli Fixing}

Let us in this section briefly summarize the results of
\cite{kklt} which are relevant for our discussion.

\subsection{GKP Compactifications}

The authors of ref.~\cite{kklt} use the GKP \cite{gkp} vacua of
the Type IIB string, which are compactified in the presence of
D-branes and orientifold planes in such a way as to preserve $N=1$
supersymmetry in four dimensions (see also \cite{sethi} for earlier discussions.). These vacua have RR and NS-NS
antisymmetric 3-form field strengths, $H_3$ and $F_3$
respectively, which can have a (quantized) flux on 3-cycles of the
compactification manifold,
\beqa \frac{1}{4\pi^2 \alpha'} \int_A F_3\ & = & J\,,\cr
 \frac{1}{4\pi^2 \alpha'} \int_B H_3\ & = & -K\,,
\eeqa
where $K$ and $J$ are arbitrary integers and $A$ and $B$ are the
different 3-cycles of the internal Calabi-Yau manifold.

The 10D field equations imply that the inclusion of fluxes of RR
and/or NS-NS forms in the compact space warps the 4D metric
according to
\be
    d\hat{s}_{10}^2 = A \, z(y) \, g_{\mu\nu}(x) \, dx^\mu \, dx^\nu
    + L^2 \, z^{-1}(y) h_{mn}(y) \, dy^m \, dy^n \,,
\ee
with a warp factor, $z$, which can be computed in regions close to
a conifold singularity of the Calabi-Yau manifold. (The
$L$-dependent factor $A$ is chosen to ensure the low-energy theory
is obtained in the 4D Einstein frame.) The result for the warp
factor is exponentially suppressed at the throat's tip, depending
on the fluxes as:
\beq z_0 \sim L^2 \, e^{-2\pi K/3g_s J} \, , \eeq
where $g_s$ is the string coupling constant. If this warp factor
suppresses standard-model particle masses relative to the string
scale, then such fluxes can naturally generate a large hierarchy
\cite{rs}.

The fluxes turned on in the GKP construction are also useful for
fixing string moduli, since they can stabilize all of those moduli
that are associated with the complex structure of the underlying
Calabi-Yau space. This includes in particular the axion-dilaton
chiral scalar field of type IIB theory, $S= e^\phi +i{\hat a}$.
From the point of view of the 4 dimensional field theory the
fluxes generate a superpotential in the effective supergravity
action of the Gukov-Vafa-Witten form \cite{gvw}:
\beq W\ =\ \int_M G_3\wedge \Omega\,, \eeq
where $G_3=F_3-iS H_3$ with $S$ the dilaton field and $\Omega$ the
unique $(3,0)$ form of the corresponding Calabi-Yau space.

This mechanism does not fix any of the moduli associated with the
K\"ahler class. The simplest models therefore only have one
modulus, which is the model-independent K\"ahler-structure modulus
containing the breathing mode, $L$, which all Calabi-Yau spaces
must have. Four-dimensional supersymmetry organizes this mode into
the complex combination $T = \hf \ReT +i \ImT$, where $\ReT
\propto L^4$ and $\ImT$ is an axion field coming from the RR
4-form, $C_4$ ($T=i\rho$ in the conventions of \cite{gkp,kklt}).
If this is the only K\"ahler modulus then it is the only one which
cannot be fixed by the fluxes.

Semiclassical dimensional reduction \cite{truncation} leads to a
K\"ahler potential of the low-energy 4D theory having the no-scale
form \cite{noscale},
\beq K\ =\ \tilde K (\varphi_i, \varphi_i^*) -3 \log \left( T+
T^*\right)\,, \eeq
with $\tilde K$ being the K\"ahler potential for all the other
fields $\varphi_i$ besides $T$. This form implies that the
supersymmetric scalar potential becomes %%FF added some bars on indices
\beq V_{SUSY}\ =\ e^K \left(\tilde K^{i\bar {\j}} D_iW
\bar{D}_{\bar {\j}}{\bar W}\right)\,, \eeq
where the sum is only over the $\varphi^i$, with $\tilde K^{i\bar
{\j}}$ being the inverse of the K\"ahler metric $\tilde K_{i \bar {\j}}
= \partial_i \partial_{\bar {\j}} \tilde K$ and $D_iW =\partial_i W +
W\partial_i \tilde K$ denoting the K\"ahler covariant derivative.
Since the superpotential does not depend on $T$, we see that the
superpotential generated by the fluxes generically fixes all
moduli but $T$.

In order to fix $T$ \KKLT\ first choose fluxes to obtain a vacuum
for which $W = W_0\neq 0$. %%FF modified the following phrase
This by itself would imply that
 supersymmetry is
broken by the $T$ field so long as Re~$T$ is finite, because $D_T
W = K_T W_0 = -3 W_0/(T+T^*) \ne 0$. They then consider a
nonperturbative superpotential, which could be either generated by
Euclidean D3-branes or by gaugino condensation within an unbroken
nonabelian gauge sector within one of the wrapped $N$ D7-branes
of the GKP scenario.

For instance, the gauge coupling for such a D7-brane gauge group
is ${8\pi^2}/{g_{YM}^2} = 2\pi {L^4} \langle z^{-2} \rangle_4
/{g_s}$, where
\be
    \langle z^k \rangle_4 = \int d^4 y \; \sqrt{{\det}_4 h} \; z^k \, ,
\ee
denotes the integral of a power of the warp factor over the
4-dimensional wrapped internal world volume of the relevant D7
brane. Normalizing $T$ so that $4\pi/g_{YM}^2 = \hbox{Re}\,T$
implies that the gauge-coupling function for this gauge group in
the low energy 4D supergravity is simply related to the breathing
mode: $f_{ab} = T \, \delta_{ab}$. Well-established arguments
\cite{gauginocondensation,ourgc} then imply the effective theory
below the gaugino-condensation scale has a nonperturbative
superpotential of the form $W_{np}= A e^{-aT}$, for appropriate
constants $A$ and $a$.

Combining the two sources of superpotentials
\beq
 W= W_0 + A e^{-aT},
\eeq
gives an effective scalar potential for the field $T = \hf \, \ReT
+i \ImT$ of the form
\beq
 V_F  =  \frac1{8 \ReT^3} \left\{ \frac13|2\ReT W^\prime
 - 3 W|^2 - 3|W|^2 \right \}\,, \label{spot}
\eeq
where $W^\prime$ denotes $dW/dT$ and $\ReT\equiv 2{\rm Re} T$.
%%FF introduced definition of ReT to be more explicit
 This has a nontrivial minimum
at finite $T$, as well as the standard runaway behaviour towards
infinity. The nontrivial minimum corresponds to negative
cosmological constant and gives rise to a supersymmetric AdS
vacuum. (More general superpotentials have also been considered in
\cite{egq}.)

\subsection{Anti-Branes and Supersymmetry Breaking}

In order to obtain a de Sitter vacuum, \KKLT\ introduce anti-D3
branes, and in so doing break the supersymmetry of the underlying
GKP vacuum. As a result the low-energy Lagrangian contains two
very different kinds of terms: those which can be organized into a
standard 4D $N=1$ supergravity Lagrangian, and those which
cannot.\footnote{In reference \cite{bkq} the effect of the
antibranes was achieved by adding fluxes of magnetic fields on the
D7 branes, in such a way that supersymmetry breaking can be made
parametrically small compared with the string scale. Consequently
the effective field theory realizes supersymmetry linearly, with
supersymmetry spontaneously broken by a Fayet-Iliopoulos term.
Ref.~\cite{eva} accomplishes a similar end using a different local
minimum of the potential for the complex structure moduli.
See also \cite{Brustein:2004xn} for interesting related discussions.} The
nonsupersymmetric terms can arise in the effective theory even
though the underlying theory is fully supersymmetric, to the
extent that the energy scale defining the low-energy theory is
smaller than the mass splittings within some supermultiplets. In
this case some of the light fields no longer have superpartners
within the low-energy theory, which can therefore only nonlinearly
realize supersymmetry \cite{nonlinsusy}. This is the generic
situation to study brane-antibrane inflation, since supersymmetry is
broken on the branes at the string scale.

Semiclassically, the presence of anti-branes %%FF slightly modified
%%this phrase
 has the effect of adding an extra
nonsupersymmetric term to the effective 4D scalar potential of
the form,
\beq
 V_T \ =\ \frac{\hat{k}}{\ReT^3} =\ \frac{k}{\ReT^2} \,,
\eeq
%
%C: I changed the power of z_b below to 2 from 4.
%
effectively due to the tension of the anti-D3 brane. The constant
$\hat{k} = 2 z_b^2 T_3/g_s^4$ parameterizes the scale of
supersymmetry breaking in the effective 4D potential, where $z_b$
is the warp factor at the location of the anti-D3 brane and $T_3$
is the anti-brane tension. Because of the warping the anti-D3
brane energetically prefers to sit at the throat's tip, and so
$z_b = z_0$. (Because $z_0$ itself depends on $L$ --- and so also
$\ReT$ --- it is convenient to follow ref.~\cite{kklmmt} and
extract these factors and so replace $\hat{k}$ with the {\it bona
fide} constant $k$.) This addition to the potential has the
effect, for suitable values of $k$, of lifting the original
anti-de Sitter minimum to a de Sitter one.

\subsection{Seeking Slow Rolls}

In the cosmological models of \cite{kklmmt} the above construction
is supplemented with a D3 brane which was free to move and so
whose position modulus, $\phi^i$, appears in the low-energy
theory. The appearance of this modulus in the K\"ahler function
becomes \cite{Giddings}
\beq
 K(T,T^*,\phi,\phi^*) = - 3 \log r \,,
\eeq
with $r = T + T^* - k(\phi,\phi^*)$ where $k(\phi,\phi^*)$ is the
K\"ahler function of the underlying Calabi-Yau space.

The interaction between the mobile D3-brane and the anti-D3-brane
also introduces another kind of nonsupersymmetric energy into the
effective potential. This potential has the form \cite{kklmmt}
\beq
 V_{int} = \frac{\hat{k}'}{r^3} \, G(\phi,\phi_0) = \frac{k'}{r^n} \,
 G(\phi,\phi_0) \, ,
\eeq
where $G(\phi,\phi_0)$ is the solution to $\nabla^2 G = \delta$ in
the background geometry, where the $\delta$ function is located at
the position, $\phi_0$, of the antibrane. Here again $\hat{k}'
\propto z_b^2 \, \hat{z}_b^2 \, T_3/g_s^4$, where as before $z_b$
is the warp factor at the position of the anti-brane and
$\hat{z}_b$ is the warp factor at the mobile brane's position. The
second equality above extracts from $\hat{k}'$ the $\ReT =
\hbox{Re}\, T$ dependence which is implicit in these warp factors.

There are two regimes for which the function $G$ is known
explicitly. Firstly, for small proper separation,
$s(\phi,\phi_0)$, between the mobile D3-brane and the anti-D3
brane this propagator varies approximately as $s^{-4}$.
Alternatively, $G$ is also known within the throat, since there
coordinates may be chosen for which the metric is given
approximately by
\beq \label{throatmetric}
    ds^2 = h^{-1/2}(r) \eta_{\mu\nu} dx^\mu dx^\nu + h^{1/2}(r)
    \Bigl[ dr^2 + r^2 ds_5^2 \Bigr] \,,
\eeq
where we do not require the explicit form of the warp factor,
$h(r)$. Within this throat region the function $G$ is related to
the corresponding quantity for anti-de Sitter space, which is
known quite generally in $n$ dimensions \cite{BL}. For anti-branes
located at the throat's tip ($r=0$) and for mobile branes
separated from this purely in the $r$ direction the result is
given explicitly by $G(r) = C_1 + C_2/r^4$, where $C_1$ and $C_2$
are constants.

The above expression assumes the antibrane is at the throat's tip,
and the power $n$ is determined by the warping at the position of
the mobile D3 brane. It is given by $n = 1$ if the mobile brane is
also near the throat's tip, or $n = 2$ if the mobile brane is not
deep inside the throat. We find that our later results do not
depend strongly on the value taken for $n$, and we choose $n=2$
for the numerical work described in later sections.

It is with the above model (with $n=2$) that the authors of
ref.~\cite{kklmmt} unsuccessfully sought solutions for which the
motion of the D3 brane was sufficiently slow to allow an
inflationary slow roll. See also refs.~\cite{hkp,henry,Buchel:2003qj,riotto} for
subsequent proposals within a similar framework.

\subsection{Sticking the Standard Model in the Throat}

Let us now briefly describe how the \KKLT\ scenario can be
extended in order to include chiral matter on the anti D3 brane,
following the recent discussion of \cite{cmqu}. The idea is to
generalise the geometry in such a way that a $\IZ_N$ singularity
can be located somewhere within the Calabi-Yau geometry, such as
at the tip of the throat where the anti D3 brane lives. This has
the end result that the throat is locally like a complex plane and
at the tip of the throat there is a 4-torus (coming from a double
elliptic fibration). Modding out by the discrete symmetry $\IZ_N$
induces fixed points on the 4-torus. The standard model (anti)
D3-branes can then be located at one of these fixed points. The
cancellation of RR tadpoles then implies that there must also be
D7 branes wrapped on the 4-torus at the throat's tip, as well as
further D3 or anti D3 branes positioned at the other fixed points
to cancel the tadpoles at each fixed point. (See Figure
\ref{figthroat} for a cartoon of the underlying geometry.)

\EPSFIGURE[ht]{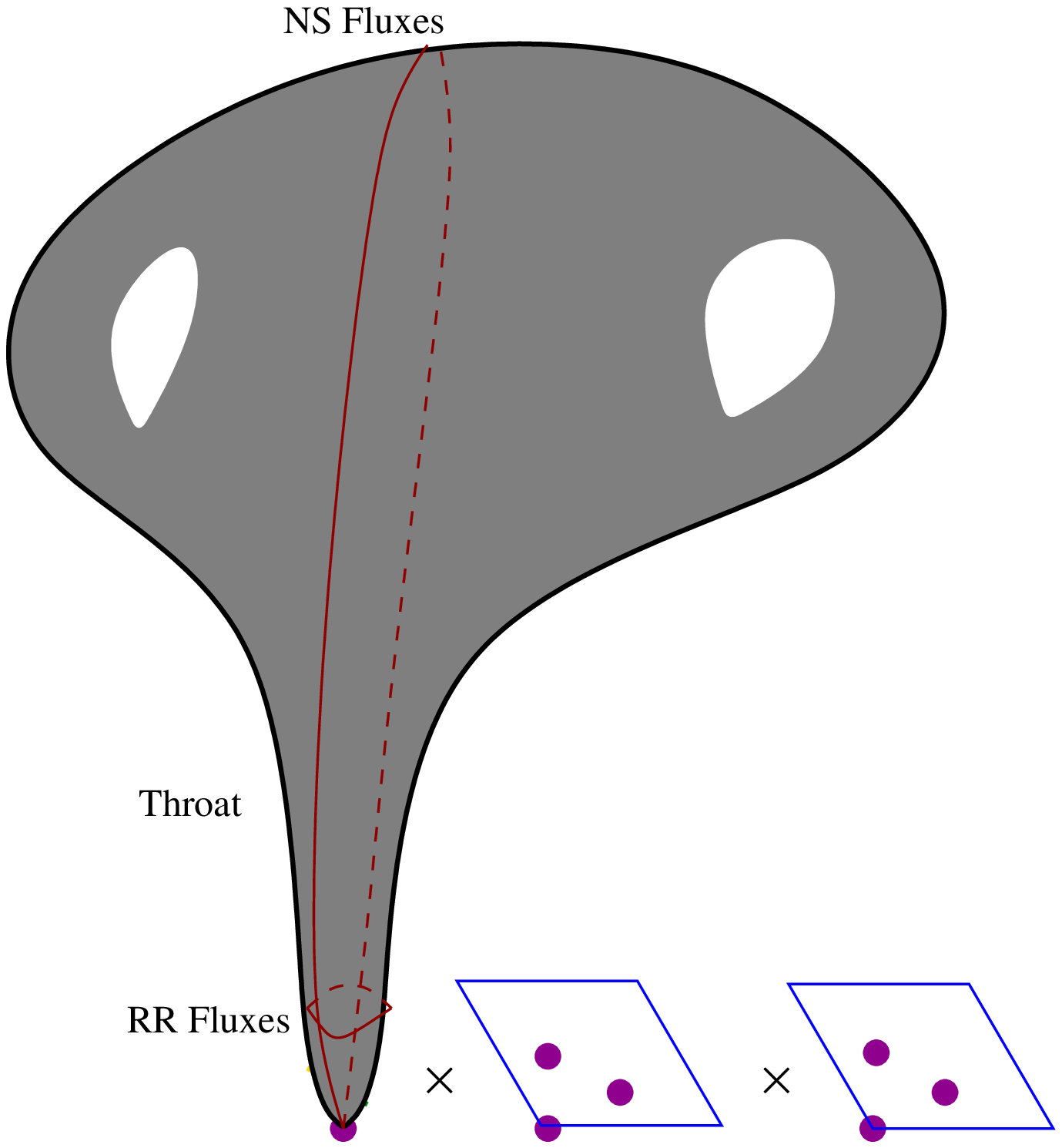,width=4in}{Description of a warped
throat with 3-form fluxes embedded in a compact geometry, with
anti-D3-branes trapped at the tip of the throat. At the tip of the
throat there is also a twisted $T^4$ with the 9 fixed points under
a $\IZ_3$ symmetry indicated. D7 branes can wrap 4-cycles outside
the throat and/or the $T^4$ in the tip of the throat, (anti) D3
branes live at the fixed points. \label{figthroat}}

The presence of these fixed points has several consequences.
First, the various tadpole conditions imply the existence of
chiral fields and gauge interactions on the branes localized at
the fixed-points, and these can contain gauge groups and matter
fields which contain the Standard Model spectrum \cite{smbranes}.
They also potentially introduce new moduli corresponding to the
blowing-up mode, $\ReB$, for each of the corresponding
singularities. Therefore, even if we follow \KKLT\ by starting
with a model with all complex structure moduli fixed by fluxes and
having only a single K\"ahler modulus which is fixed by
nonperturbative effects associated with D7 branes away from the
throat, having new fixed points introduces new moduli whose
stabilization must be addressed.

As it happens, these new moduli generally do appear in the
low-energy supergravity potential, in the form of a $D$-term. They
typically do so because supersymmetry pairs the blowing-up modes
with axion fields, $\ImB$, which are required to cancel the
anomalies which the chiral fermions of the low-energy theory have
for various $U(1)$ gauge groups. To see why this is so, notice
that anomaly cancellation requires the axion field to have both of
the couplings $\partial_\mu \ImB \, A^\mu$ and $\ImB \, F_{\mu\nu}
\widetilde F^{\mu\nu}$, where $F = dA$ is the gauge field for the
anomalous $U(1)$. But supersymmetry in 4D then relates these two
couplings to others involving the blowing up modes. In terms of
the chiral scalar, $B = \ReB + i \ImB$, containing both the axion
and the blowing up mode, anomaly cancellation requires the
following two couplings \cite{dsw}.\footnote{See \cite{bdkv} for a
recent discussion of Fayet-Iliopoulos terms in a more recent
context.}

\begin{itemize}
\item
The coupling $\ImB \, F_{\mu\nu} \widetilde F^{\mu\nu}$ requires
the holomorphic gauge coupling function must depend on $B$,
according to
\beq\label{gaugecoupling} f=f_0 +\alpha B \eeq
for some nonzero constant $\alpha$. The form for $f_0$ depends on
the origin of the anomalous $U(1)$. If it is associated with a D7
brane which lies far from the throat, then $f_0 = T$ as was the
case for the gaugino condensation within the \KKLT\ framework. On
the other hand, if the anomalous $U(1)$ arises from a D7 brane
located at the tip of the throat then the gauge coupling function
goes like $L^4/z_0^2$, and so the warping cancels the $L$
dependence, leading to a $T$-independent result: $f_0 = S = $
(constant). A $T$-independent gauge coupling function is also what
is expected if the anomalous $U(1)$ is associated with a D3 brane
situated anywhere within the internal 6 dimensions.

\item
The $\partial_\mu \ImB A^\mu$ term requires the chiral scalar $B$
can only appear in the K\"ahler potential together with the
anomalous $U(1)$ gauge multiplet, $A$, through the combination $B
+ B^* + A$. This in turn implies the existence of a
Fayet-Iliopoulos $D$-term \cite{fi} proportional to $\partial
K/\partial B$.
\end{itemize}

Using these in the 4D supergravity action (and expanding the
K\"ahler function to leading order in powers of the chiral matter
fields, $Q$) leads to a contribution to the low-energy 4D scalar
potential for $B$ having the $D$-term form:
\beq
 V_D = g^2\left[ \left( \frac{\partial K}{\partial B}
 \right)_{A=0} - Q^\dagger q Q\right]^2 \, ,
\eeq
where $g^{-2}= f+f^*$ is the inverse 4D gauge coupling for the
anomalous $U(1)$. Here $q$ generically denotes the charge matrix
of the chiral matter fields under the anomalous $U(1)$. Clearly
this potential generically lifts the flat directions associated
with $B$, as does gaugino condensation by any nonabelian
gauge-group factors for which the gauge coupling function is given
by eq.~(\ref{gaugecoupling}).

\section{The Effective Theory}

With an eye to searching for inflation, in this section we write
down a four dimensional effective field theory which is meant to
capture the low-energy dynamics of a mobile D3 brane moving within
the string vacua discussed above (and hopefully for more general
configurations). To this end we include in the low energy theory
representatives of each of the types of moduli described in the
previous section as well as the position modulus of the mobile D3
brane. Our goal is to follow the way that the mobile brane and
other moduli move under the influence of the geometry and forces
due to the other branes.

Thus, we choose a Lagrangian which depends on the following
fields: (1) the moduli $T$ and $B$; (2) various gauge multiplets,
including at least one anomalous $U(1)$ multiplet, $A$; (3) the
chiral matter fields, $Q$, whose fermions are responsible for the
$U(1)$ anomalies; and (4) the position modulus, $\phi$, of the D3
brane whose motion is the putative inflaton. If we were to set the
fields $A$, $B$ and $Q$ to zero, we would be left with a
Lagrangian only for the fields $T$ and $\phi$ for which the
analysis of \cite{kklmmt} applies (and shows that inflation is not
easily generated). In this sense our scenario generalises the one
in \cite{kklmmt}. For simplicity of analysis we specialize to the
case of a single anomalous $U(1)$ multiplet, $A$, to a single
charged field, $Q$, and to a single modulus, $B$, but we expect
our results to also apply to the more realistic case where several
such moduli appear.

We next discuss in turn the supersymmetric and nonsupersymmetric
contributions to the low-energy scalar potential.

\subsection{Supersymmetric Terms}

We imagine the gauge group of the low-energy theory to include
unbroken nonabelian factors associated with some of the D7 and D3
branes of the model, and that these gauge interactions confine at
energies below the compactification scale. We further imagine some
of the relevant D7 branes are located far from the throat, and so
have gauge coupling functions ${\cal F}_{ab} = ({\cal F}_0 + T)
\delta_{ab}$. We also assume there to be a nonabelian gauge group
associated with the branes which are localized at the orbifold
points, and that these are located at the throat's tip and so have
gauge coupling functions of the form ${\cal F}^t_{ab} = ({\cal
F}^t_0 + B) \delta_{ab}$. As discussed above, in these expressions
${\cal F}_0$ and ${\cal F}^t_0$ are constants which are
independent of the low-energy moduli of interest.

These nonabelian gauge groups are assumed to exist in addition to
the anomalous $U(1)$ gauge group mentioned above, whose gauge
fields survive into the low-energy theory. Since this $U(1)$ is
assumed to be associated with a brane at the throat's tip, its
low-energy gauge coupling function, $f$, is independent of $T$.
$f$ must depend on $B$, however, since this is required by the
condition that Im$B$ cancel the $U(1)$ anomalies in the low-energy
theory.

Since our interest is in the effective theory below the
condensation scale for the nonabelian gauge groups, their gauge
degrees of freedom can be integrated out. Their sole low-energy
influence is then through the nonperturbative effects to which
they give rise (like gaugino condensation
\cite{gauginocondensation,ourgc}) since these generate
contributions to the low-energy superpotential of the form $W_{np}
= A \, e^{-a T} + C \, Q^\ell \, e^{-c  B}$. (The power, $\ell$,
of $Q$ which appears here is determined by the condition that
$W_{np}$ be invariant under the gauged $U(1)$ symmetry, under
which Im$B$ shifts like a would-be Goldstone boson.) In general
the quantities $A$ and $C$ could also depend on other moduli, such
as $\phi$ \cite{kklmmt,henry}, but this dependence need not be
strong, particularly if the mobile D3 brane should be distant from
the brane on which the condensation occurs. We do not use this
$\phi$ dependence in what follows.

We are led in this way to describing the supersymmetric part of
the low-energy theory (below the confinement scale of the
nonabelian gauge bosons) by a 4D supergravity model which is
characterized by the following K\"ahler function, $K$,
superpotential, $W$, and $U(1)$ gauge coupling function, $f$:
\bea \label{KfandW}
    K &=& -3 \, \ln r + \cB \left( B + B^* + A \right) + {\cQ} \left(
    Q^\dagger \, e^{A} , Q \right) \nonumber \\
    f &=& f_0 + \alpha \, B
    \qquad \hbox{and} \qquad W = w + W_p(Q) + W_{np}(T,B) \,.
\eea
Here $\cB$ and $\cQ$ are arbitrary real functions of their
arguments, and $w$ is the term in the superpotential which arises
in the low-energy theory due to the fluxes which fix the
complex-structure moduli. We follow \KKLMMT\ by taking $w$ to be
independent of $T$, $\phi$ and $B$, and by taking $r = T + T^* -
k(\phi^*,\phi)$ \cite{Giddings,kklmmt}, with the logarithm
expressing the leading dependence of $K$ on $T$ and $\phi$ for
large $r$. The function $k(\phi,\phi^*)$ is the K\"ahler potential
for the underlying metric on the internal Calabi-Yau space itself.
As discussed above, the quantity $f_0$ is a constant if the
anomalous $U(1)$ is associated with a D3 brane, or for a D7 brane
localized at the throat's tip. For D7's located elsewhere we
instead have $f_0$ proportional to $T$ which was used to induce the
nonperturbative dependence on $T$ of $W$.
%%FF added previous phrase
 Finally, we include a
perturbative superpotential, $W_p(Q)$, since in general charged
matter fields acquire cubic tree-level superpotentials in the
low-energy theory.

The contributions of $r$ and $B$ to the K\"ahler function arise as
a sum of contributions as written above since microscopically the
$B$'s are associated with fixed points while $T$ and $\phi$ are
not. (A similar split would also have occurred if $B$ had been an
unfixed complex-structure modulus.) Similar remarks apply to the
dependence of $K$ on $Q$, but this does not significantly
influence our results. With these choices the kinetic energies for
the fields are controlled by the following K\"ahler metric
\bea \label{kahlermetric}
    ds^2 &=& \frac{3}{r^2} \Bigl[ dT^* dT + \left( r \,
    k_{\phi\phi^*} + k_\phi \, k_{\phi^*} \right) d\phi^* d\phi -
    \left( k_\phi \, dT^* d\phi + \hbox{c.c.} \right) \Bigr] \nonumber\\
    &&\qquad \qquad \qquad + \cB'' \, dB^* dB + \cQ_{Q{Q}^*}
    \, d{Q}^* dQ
    \,,
\eea
where for simplicity we restrict to a single charged chiral field,
$Q$. As usual subscripts denote differentiation with respect to
the appropriate fields, and primes denote differentiation with
respect to the function's argument.

The $D$-term and $F$-term contributions to the scalar potential $V
= V_F + V_D$ for such a supergravity become
\be  \label{VD}
    V_D = \frac{( \cB' + q\cQ_{\overline{Q}} Q)^2 }{f + f^*} =
    \frac{( \cB' + q \cQ_{\overline{Q}} Q )^2}{f_0 + f_0^* +
    \alpha (B + B^*) }  \,,
\ee
and
\bea \label{VF}
    V_F &=& \frac{e^{\cB + \cQ}}{r^3} \left[ \frac{r^2  \left|
    W_T \right|^2 }{3} \left(1 + \frac{ k_\phi k_{\phi^*}}{r
    k_{\phi\phi^*}} \right)  - r \, \Bigl( W^*
    W_T + \hbox{c.c.} \Bigr) \right. \\
%    && \qquad\qquad  +\frac{r\left|W_\phi\right|^2 +
%    r\left(k_\phi W_T W^{*}_\phi+ \hbox{c.c.}\right)}{3 k_{TT^*}}
%    \nonumber \\
    && \qquad\qquad + \left. \frac{ \left|W_B+\cB' \, W
    \right|^2 }{\cB''} + \frac{\left|W_Q + \cQ_Q \, W \right|^2}{\cQ_{Q\overline{Q}}}
    \right] \,. \nonumber
\eea

\subsection{Supersymmetry-Breaking Terms}

Because the realistic models typically involve both branes and
antibranes, for the antibranes supersymmetry is broken at the
scale of the brane tension ({\it i.e.} the string scale, modified
by the warp factor appropriate to the brane position).
Consequently, in addition to the previous supergravity Lagrangian
we must also include in the low-energy theory terms which
explicitly break supersymmetry.\footnote{As mentioned earlier,
strictly speaking these terms nonlinearly realize supersymmetry.
But experience with gauge theories \cite{NLvsExpl} indicates that
this is likely to be indistinguishable in a unitary gauge from
explicit breaking.}

Although a general statement of the low-energy form of the
Lagrangian for branes interacting with antibranes is not yet known
\cite{nonlinsusy}, approximate expressions may be obtained in the
limit that the antibrane supersymmetry-breaking effects are
perturbatively small.\footnote{See \cite{ciu} for a recent
discussion of soft-breaking terms induced by fluxes and by
antibranes.} In this limit their leading effects may be added to
the supergravity Lagrangian considered above. We obtain then the
low-energy 4D scalar potential $V= V_F + V_D + V_{sb}$ with $V_F$
and $V_D$ given as above, and where $V_{sb}$ can also be written
as a sum of two terms: the antibrane's tension term plus a
brane-antibrane interaction term, $V_{sb} = V_T + V_{int}$. We
expect the supersymmetry breaking also introduces the soft
susy-breaking terms, $V_{soft}$, including trilinear terms and
scalar masses for matter fields, which the low-energy
phenomenology of the $Q$ fields would require.

The tension part of the supersymmetry-breaking potential takes the
form
\be \label{VT}
    V_{T} = \frac{k}{r^2} \,,
\ee
which is the \KKLT\ result for antibranes localized near the tip
of the throat. Recall that if the supersymmetry breaking is due to
anti-D3 branes, then $k$ is related to the brane tension $T_3$ by
$k = 2 z_b^2 T_3/g_s^4$.

As before, the brane-antibrane interaction term similarly is given
by
\be
\label{Vint}
    V_{int} = \frac{k'}{r^2} \, G(\phi,\phi_0)
\ee
where $\phi_0$ denotes the coordinate position of the antibrane.
In what follows we shall use the explicit expression
\beq \label{Glikepsi-4}
    G \propto \frac{1}{(\ModPhi - \ModPhi_0)^{4}} \,,
\eeq
with the real coordinate $\ModPhi$ representing the position
of the mobile brane and $\ModPhi_0$ denoting the antibrane
position in the same coordinates. This expression follows quite
generally if the proper separation between the brane and antibrane
is very small compared with the geometry's radius of curvature,
since in this case $G \propto s^{-4}$ and $s^2 \approx \sum_i
|\phi^i - \phi_0^i|^2$. The above expression then follows provided
we choose $\ModPhi^2 = \sum_i |\phi^i|^2$, and minimize over the
directional variables of $\phi^i$.

Alternatively, eq.~\pref{Glikepsi-4} also follows if the
antibranes are at the throat's tip, with the mobile brane also
within the throat. This follows if we work in the coordinates
defined by the metric of eq.~\pref{throatmetric}, and choose
$\ModPhi = \ModPhi_0 + r$.

\section{Inflationary Dynamics}

In this section we explore the dynamics which is implied by the
above 4D effective theory, with the goal of identifying the
circumstances under which it permits an inflationary slow roll,
given that all of the potential moduli are left free to move. We
do find that inflation is generically difficult to obtain (in
agreement with \cite{kklmmt}), and we find that this property is
not significantly changed by the presence of new moduli like $B$
in our model. Unlike these authors we explicitly identify some of
the special circumstances which do allow inflation, in order to
see how unusual inflationary solutions are. We also identify in
this section the observational consequences which follow from the
inflationary trajectories which we do find.

\subsection{Domain of Validity of Approximations}

Before presenting numerical results, we briefly pause to describe
some preliminary considerations which are useful to have in mind
when performing the detailed analysis. Because our scalar fields
have nonminimal kinetic terms, the scalar parts of our low-energy
effective actions have the generic form
\be \label{generalaction}
    {S}_{\rm scalar} = - \int d^4x \; \sqrt{-g}
     \left[ \, \frac{1}{2} \, g_{ij} \,
    \partial_\mu \varphi^i \,
    \partial^\mu \varphi^j + V(\varphi) \right] \,,
\ee
with the target-space metric, $g_{ij}$, given by
eq.~\pref{kahlermetric} and the scalar potential, $V$, as
discussed at length in previous sections. The field equations
which follow from this Lagrangian density are
\bea  \label{fullequations}
    \ddot \varphi^i &+& 3H \dot \varphi^i + \Gamma^i_{jk}
    \, \dot\varphi^j \, \dot \varphi^k
    + g^{ij} \, {\partial V\over\partial\varphi^j} = 0 \,,\nonumber\\
    H^2 &=& \left({\dot a\over a}\right)^2 = {8\pi G\over3} \left(\frac12
    g_{ij}\dot \varphi^i \dot \varphi^j +V\right) \,,
\eea
where $a$ is the scale factor for the 4D FRW metric, dots denote
time derivatives and the $\Gamma^k_{ij}$ denote the target-space
Christoffel symbols constructed using the metric $g_{ij}$. Notice
that the new terms involving the Christoffel symbol in the scalar
field equation are quadratic in time derivatives, and so are
normally negligible in the slow-roll approximation.\footnote{See
for instance \cite{Doug} for a recent study of some of the
possible cosmological effects of these terms.} Even so, in later
sections we integrate the full equations when numerically
computing scalar evolution.

Because of the nontrivial target-space metric we must adopt
slow-roll criteria which differ slightly from those normally used.
An invariant notion of the slowness of the scalar evolution is
controlled by the smallness of the generalized slow-roll
parameters defined by $\epsilon$ and $\eta = \hbox{min}_a \,
\eta_a$, where\footnote{We use the convention $M_p^2 = 8 \pi G$,
and take $M_p = 1$ for our numerical work.}
\be
    \epsilon = \frac{M_p^2}{2} \, g^{ij} \left( \frac{ V_i \, V_j}{V^2}
    \right) \qquad \hbox{and} \qquad
    {N_i}^j v^{(a)}_j = \eta_a \, v^{(a)}_i \, .
\ee
Here $V_i = \partial_i V = \partial V/\partial \varphi^i$ and the
$\eta_a$ are eigenvalues of the matrix, ${N_i}^j = M_p^2 \, g^{jk}
\, V_{;ik}/V$. The target-space covariant derivative is built in
the usual way from the target-space Christoffel symbols: $V_{;ij}
= \partial_i\partial_j V - \Gamma^k_{ij} \partial_k V$. These
expressions agree with the usual ones when $g_{ij} = \delta_{ij}$,
and are invariant under scalar field redefinitions.

A second general consideration involves the domain of validity of
the low-energy effective theory. In particular, the derivation of
the form of the 4D Lagrangian for the moduli presupposes the
internal dimensions to be much larger than the string scale. In 4D
Planck units this requires $r, \ReT \gg 1$. Since it is more
convenient numerically to deal with fields which are $O(1)$, many
of the solutions we find also have $r \sim \ReT \sim O(1)$. We
therefore return at the end of the next section to show how to
construct solutions having larger values of $r$ and $\ReT$ from
the ones we present numerically.

\subsection{Inflationary Dynamics}

We now describe our numerical results which explore the
inflationary possibilities of our effective 4D theory in more
detail. These results were obtained by numerically evolving the
scale factor, $a$, of the 4D metric and {\it all} of the scalar
moduli forward in time, starting from rest, using the full
equations, \pref{fullequations}. In practice this is most easily
done by using the scale factor itself as the time variable when
integrating the scalar field equations.

The scalar fields whose evolution we follow are the complex
quantities $T = \hf \, \ReT +i \ImT$, $B = \ReB + i \ImB$, $Q$ and
$\phi^i$, with $i = 1,2,3$. The numerical evolution we describe
was performed using the target-space metric defined by
eq.~\pref{kahlermetric}, such as follows for the scalar part of
the 4D supergravity defined by the functions $K$, $f$ and $W$ of
eqs.~\pref{KfandW}. For simplicity we take the function ${\cal B}$
to be ${\cal B} = \hf \, (B + B^*)^2$, and approximate the
Calabi-Yau K\"ahler function by $k(\phi^*,\phi) = \sum_i
|\phi^i|^2$. We take the scalar potential to be the sum of $V_D$,
$V_F$, $V_T$ and $V_{\rm int}$, as given by eqs.~\pref{VD},
\pref{VF}, \pref{VT} and \pref{Vint}.

\subsubsection{Qualitative Description}

Although our full numerical simulations follow all of the moduli,
for the inflationary solutions we present, none of the fields are
very important besides the two fields $\ReT = 2 \hbox{Re}\, T$ and
the brane position, $\ModPhi^2 = \sum |\phi^i|^2$, (whose relation
to coordinates on the Calabi-Yau space is given below
eq.~\pref{Glikepsi-4}).
%%FF added a factor of 2 in definition of ReT above and closed the
%%parenthesis
 This is because for these solutions the
other fields tend to roll quickly to the local $\ReT$- and
$\ModPhi$-dependent minima of their potentials, leaving the
long-term behaviour dominated by the $\ReT$ and $\ModPhi$
evolution. It is therefore instructive to approximately integrate
out these other fields analytically in order to understand
qualitatively the potential which governs the inflationary
dynamics. This is the topic of the present section.

The field $Q$ %%FF erased a very
 generally likes to very quickly sit at its
minimum, and so does not contribute to the scalar dynamics in an
interesting way. In what follows all we need to know about this
minimum is that it occurs at nonzero $Q$. This would automatically
be preferred if the power of $Q$ in the superpotential $Q^\ell
\,e^{-cB}$ were to satisfy $\ell \le 1$. The stabilization of $Q$
away from zero is automatic if $\ell < 1$ since in this case
$\partial W/\partial Q$ involves negative powers of $Q$, which
drive the field away from zero.\footnote{We find numerically that
the case $\ell = 1$ also turns out to stabilize $Q$ away from
zero.} It is then stabilized at a finite value because positive
powers of $Q$ appearing in the K\"aher function also prevent a
runaway to infinity. Alternatively, stabilization of $Q$ away from
zero might also be arranged by suitably designing the interactions
of the perturbative superpotential, $W_p$, or the $Q$-dependent
soft supersymmetry-breaking terms.

It is very simple to analytically integrate out the imaginary
parts of $T$ and $B$. With the assumption $k(\phi^*,\phi) = \sum_i
|\phi^i|^2 = \ModPhi^2$, only $V_{\rm int}$ cares about the
direction, $\ArgPhi$, of $\phi^i$, and this (for real $\phi_0$) is
minimized by setting all components of $\phi^i$ to zero apart from
the overall modulus $\psi$. Once this is done we have $r = \ReT -
\ModPhi^2$. It is only $V_F$ which depends on Im$T = \ImT$ and
Im$B = \ImB$, with
\bea
    V_F &=& {e^{2 \ReB^2} \over r^3}\left(
    2\,{\frac {AC \left( ra +4\,{\ReB}^{2}-2 \,c \ReB \right) \cos \left( \ImT -
    \ImB \right) }{{e^{a \ReT/2}}{e^{c \ReB}}}}+2\,{\frac {A w \left( r a+4
    \,{\ReB}^{2} \right) \cos  \ImT  }{{e^{a\ReT/2}}}} \right. \nonumber\\
    &+&4\,{
    \frac {\ReB C w \left( 2\,\ReB-c \right) \cos \ImB }{{e^{c \ReB}}}
    } \left. +{\frac {{A}^{2} \left( 12\,{\ReB}^{2}+6\,r a+r\ReT\,{a}^{2}
     \right) }{3{e^{a\ReT}}}}+{\frac {{C}^{2} \left( 2\,\ReB-c \right) ^{2}}
    { \left( {e^{c \ReB}} \right) ^{2}}}+4\,{\ReB}^{2}{w}^{2} \right) \nonumber\\
\eea
where $A$, $a$, $C$, $c$ and $w$ are the constants which appear in
the superpotential (after the vacuum value of $Q$ is absorbed),
\be
    W = w + A \, e^{-a T}  + C \, e^{-cB} \, ,
\ee
which we take to be real and positive. Minimizing with respect to
$\ImT$ and $\ImB$, we find that $w \cos\ImT = - |w|$ at the
minimum, and $\beta = 0$.

To minimize with respect to $B$, we can use the hindsight that $B$
typically evolves toward small field values, since it is only the
exponential superpotential, $C e^{-cB}$, which excludes the
solution $B = 0$. For small $B$ we can take the gauge kinetic
function to be a real constant, $f = f_0$, without loss of
generality, and so $V_D = 2|B|^2/f_0$. For small $B$, $V_F$ can be
approximated by expanding to quadratic order in $B$, giving
\be \label{intoutB}
    V(B)  \cong \alpha + \beta B + \frac12\gamma B^2 \,,
\ee
where $\alpha$, $\beta$ and $\gamma$ are calculable functions of
$r$ and $\ReT$. The value of $B$ which minimizes the potential is
then $B = -\beta/\gamma$, and the value of the potential at this
minimum is $V_{\rm min} = \alpha - {\beta^2/ 2\gamma}$. We have
verified that this actually gives a quite good approximation to
the full $B$-dependent potential once $B$ has settled down to its
instantaneous local minimum, due to the Hubble damping of its
oscillations during inflation.

\DOUBLEFIGURE[ht]{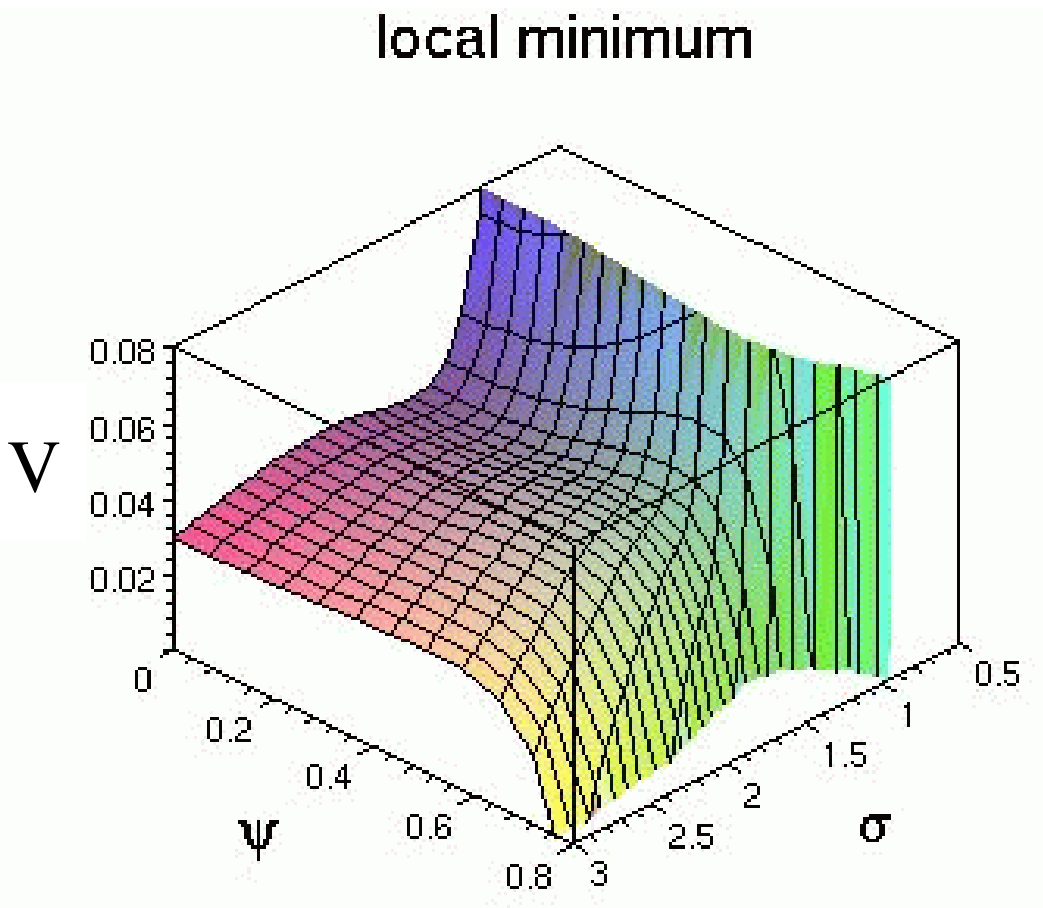,width=1.4\hsize}{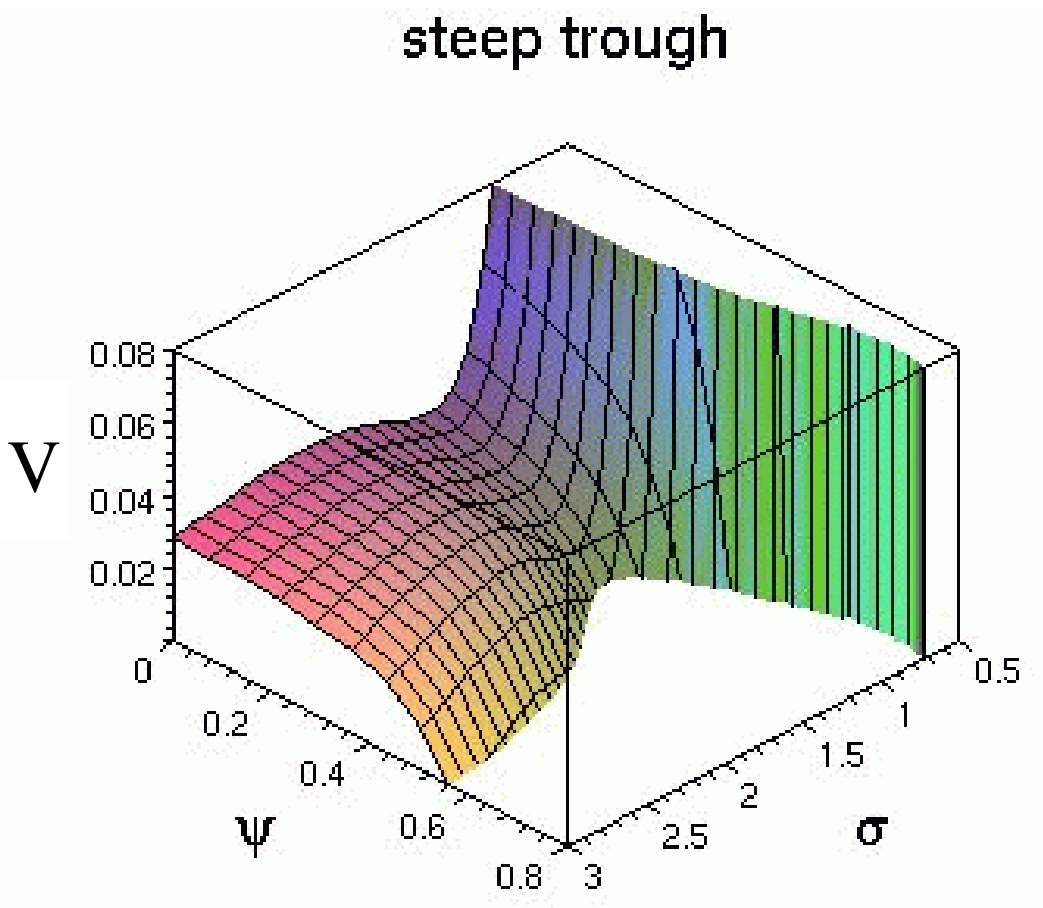,width=1.4\hsize}
{The potential with a local minimum in the trough, which leads to
old inflation.\label{figpot1}} {The potential with a steep trough,
which gives a short period of inflation.\label{figpot2}}

Given these values for the other moduli, we may now examine the
potential $V = V_D + V_F + V_{sb}$ as a function of $\ReT$ and
$\psi$ to discover the conditions that will lead to inflation,
typically with $\psi$ playing the role of the inflaton. The
parameter space for the full potential is large, consisting of
$A,a,C,c,k,k',f_0,\psi_0$, where $\psi_0 = |\phi_0|$ is the
modulus of the antibrane position (see (\ref{Vint}) and the
discussion following).

Despite the large number of parameters which can be varied, we
find essentially one situation where inflation can occur. For
certain parameter values, a trough can form along the $\psi$
direction, as shown in Figs.~\ref{figpot1} and \ref{figpot2}. This
trough can have a local minimum at small values of $\psi$, as in
Fig.~\ref{figpot1}, or else it slopes monotonically toward zero
potential, in the direction of the brane-antibrane annihilation,
as in Fig.~\ref{figpot2}. The parameters chosen to obtain these
figures are
\be
    w= 0.25,\ {f_0}=1.175,{A}= 0.9,\  a= 2.1,\ {C}= 0.51,\ c=1,\
    k= 0.3,\ {\psi_0}= 1.136
\label{params}
\ee
with $k'=0.005$ chosen for Fig.~\ref{figpot1} and $k'=0.03$ for
Fig.~\ref{figpot2}. (Recall that these values are expressed in 4D
Planck units, with $M_p = (8\pi G)^{-1/2} = 1$.)

The qualitative features of this potential can be understood as
follows:
\begin{itemize}
\item
The function $\alpha$ in (\ref{intoutB}) has the form
\be
    \alpha(\ReT,\ModPhi) = {\frac { \left( 2\,a+\ReT\,{a}^{2}/3
    \right) {{A}}^{2}}{{r}^ {2}{e^{a\ReT}}}}-{\frac { \left(
    2\,wa+2\,{C}\,a \right) {A}}{{r}^{2}{e^{a\ReT/2}}}}+{\frac
    {{{C}}^{2}{c}^{2}}{{r}^{3 }}} \,.
\ee
For the chosen parameters, $\alpha$ has a minimum as a function of
$\ReT$ over the interesting range of $\ModPhi$.  This explains the
existence of a trough in the $\ModPhi$ direction.  Furthermore,
the correction $-\beta^2/(2\gamma)$ in (\ref{intoutB}) does not
destroy this trough. The minimum with respect to $\ReT$ is a
consequence of the usual modulus stabilization due to gaugino
condensation. It is only a local minimum; for larger values of
$\ReT$, there is runaway behavior toward $\ReT\to\infty$.
\item
The behavior in the $\ModPhi$ direction depends not only on
$\alpha$ but also the SUSY-breaking terms in the potential, whose
strengths are determined by $k$ and $k'$.  For example a function
of the form $k_0 r^{-n} - k'r^{-2}(\psi-\psi_0)^{-4}$ (where the
first term represents the typical dependence on $\ModPhi$ of the
F-term contributions to the potential) can be seen to have the
observed qualitative behavior of $V(\ModPhi)$ along the trough for
fixed $\ReT$. Whether there is a saddle point along the trough is
controlled by the relative sizes of the parameters $k_0$, $k'$ and
$\psi_0$. The runaway to large $\ModPhi$ arises once the
brane-antibrane attraction dominates, and describes the approach
of these two objects in prelude to their mutual annihilation.
\end{itemize}

In the case where there is a local minimum in the trough, as shown
in Fig.~\ref{figpot1}, a de Sitter solution may be obtained along
the lines of that obtained in ref.~\cite{kklt} simply by sitting
at the local minimum, provided that the parameters are chosen to
ensure the potential is positive there.  Of course this provides
at best an example of old inflation, in which most of the universe
continues to inflate forever, and so is not a phenomenologically
attractive scenario. But it is possible to obtain inflation by
tuning away the local minimum and so flattening the trough. Since,
generically, the curvature of the saddle point which separates the
local minimum from the large-$\psi$ region is a function of all
these parameters,
\be
    \left.{\partial^2 V\over \partial\psi^2}\right|_{\rm saddle\ pt.} =
    f(A,a,C,c,k,k',f_0,\psi_0),
\ee
sufficient flatness --- {\it i.e.,} $({\partial^2 V /
\partial\psi^2})|_{s.p.} = 0$ --- can be obtained by adjusting
practically any of the parameters in $V$. This highly tuned
situation is the optimal one for inflation, since going beyond it to
negative values of $({\partial^2 V / \partial \psi^2})$ leads to a
steeper trough, and so leads to an earlier end for inflation, with
typically far less than the canonical 60 $e$-foldings of
inflation.

\subsubsection{Numerical Results}

We now report on the results of the full numerical evolution of
all of the scalar moduli, using the full equations of motion,
eqs.~\pref{fullequations}.

We have explored the sensitivity of the duration of inflation to
the parameter values, and find that in order to get 60
$e$-foldings, a tuning of 1 part in 1000 is required in any given
parameter. If we tune to only a part in 100, as might have been
naively expected to suffice, we obtain only about 30 $e$-foldings.
The situation is illustrated with respect to the parameters $k'$
and $k$, which appear in the SUSY-breaking part of the potential,
in Figs.\ \ref{figkp} and \ref{figk}. It is worth remarking that a
small number of $e$-foldings like 30 could be phenomenologically
viable if the mechanism of reheating were inefficient enough to
give a reheat temperature far below the string scale \cite{CFM}.
Such a possibility is not out of the question, since the mechanism
of reheating after brane-antibrane annihilation is unknown.  If,
for instance, the false vacuum energy were initially dumped mainly
into invisible closed-string modes, the reheat temperature in
visible radiation could naturally be small.

\DOUBLEFIGURE[ht]{kprime.eps,width=\hsize}{k.eps,width=\hsize}
{Dependence of the amount of inflation on the value of the
brane-antibrane interaction strength
$k'$.\label{figkp}}{Dependence of the amount of inflation on the
tension parameter $k$ in the SUSY-breaking part of the
potential.\label{figk}}

\DOUBLEFIGURE[ht]{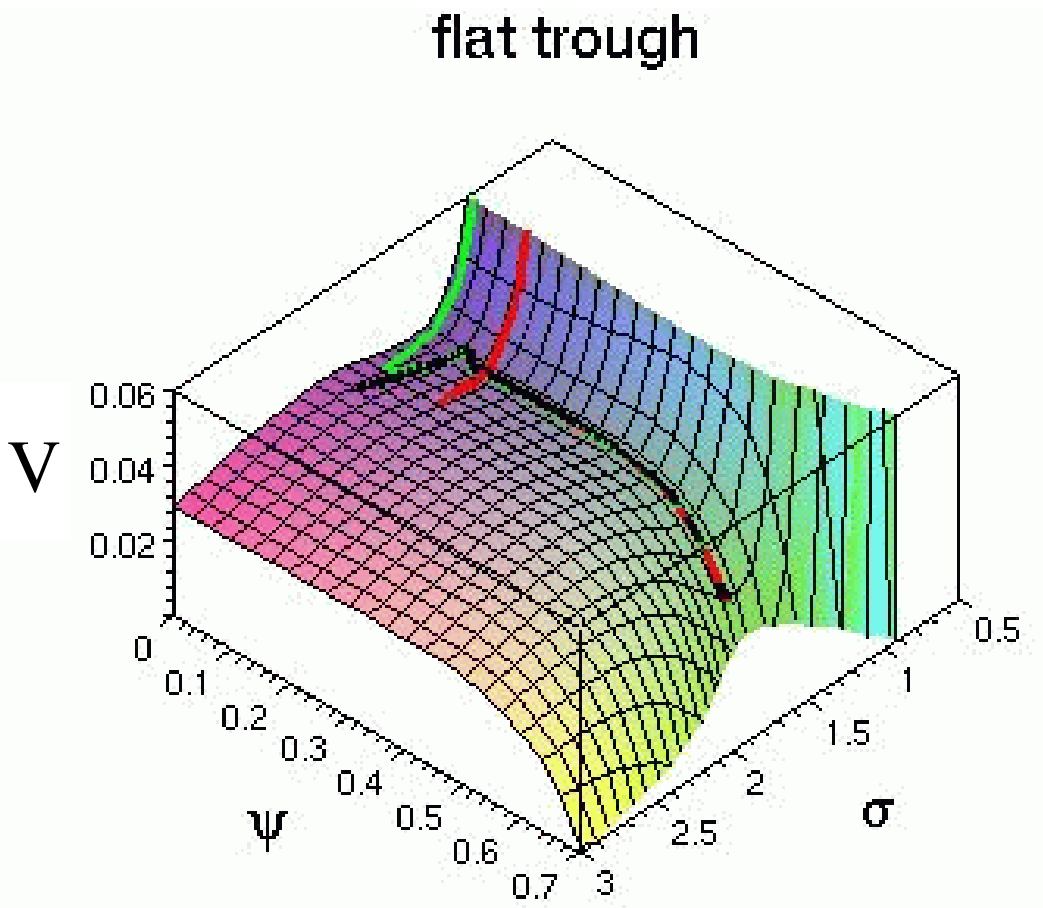,width=1.3\hsize}{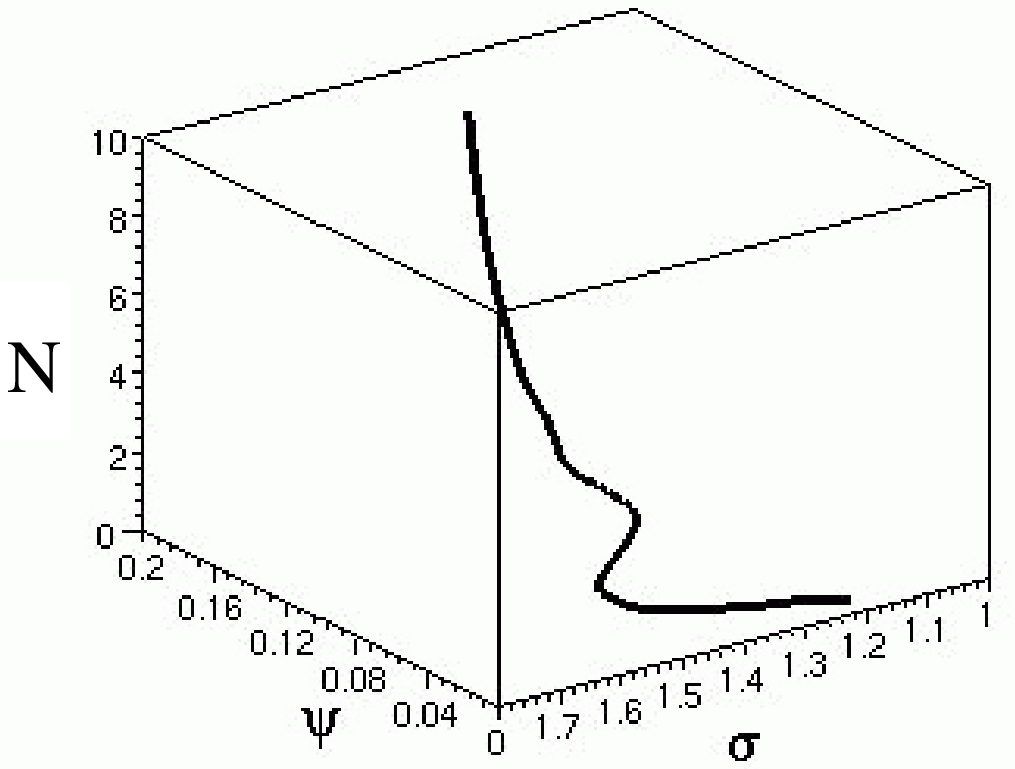,width=1.25\hsize}
{Potential with a flat trough, showing several inflaton trajectories starting
from different initial field values. Parameters are given by eq.\ (4.7) and
$k'=0.01$. \label{figtrough}}{The first 10 (out of 90)
$e$-foldings of expansion as a function of evolving field values for one of the
trajectories in Fig.\ 6.\label{figefold-curve}}

Fig.~\ref{figtrough} shows several trajectories for the fields
$\{\ReT, \ModPhi \}$, starting initially at rest, drawn on their
potential in the case where the trough is sufficiently flat to
yield up to 90 $e$-foldings of inflation. (With a finer tuning of
parameters, even more inflation is possible; this example uses the
parameters in (\ref{params}) and $k'=0.01$.) The trajectories are
integrated until the potential becomes negative; at this point the
brane and antibrane are close to annihilating, and we expect that
the low-energy effective action no longer gives an accurate
description of the system, due to large corrections at the string
scale.

In contrast to the precision of the tuning required for the
parameters of the potential in order to obtain a flat enough
potential, the sensitivity to the initial conditions of the
inflaton moving in this potential is quite mild. The fields
quickly roll to the trough, within the first $2$-$3$ $e$-foldings
of expansion, as illustrated in Fig.~\ref{figefold-curve}. The
total amount of inflation obtained is controlled mainly by the
initial value of $\psi$, which determines how much of the trough
is traversed. Fig.~\ref{figefold} shows the number of $e$-foldings
which are achieved as a function of initial field values.

For completeness, we also illustrate the evolution of $B$ during
inflation along the trough in Fig.~\ref{figB}. This figure shows
that the initial transient oscillations die out within the first 4
$e$-foldings, and $B$ remains nearly constant during the slow-roll
period. Furthermore, its small value during this roll justifies
the heuristic use of a Taylor expansion in $B$ of the full
potential, which we employed for the qualitative description given
above.

\DOUBLEFIGURE[ht]{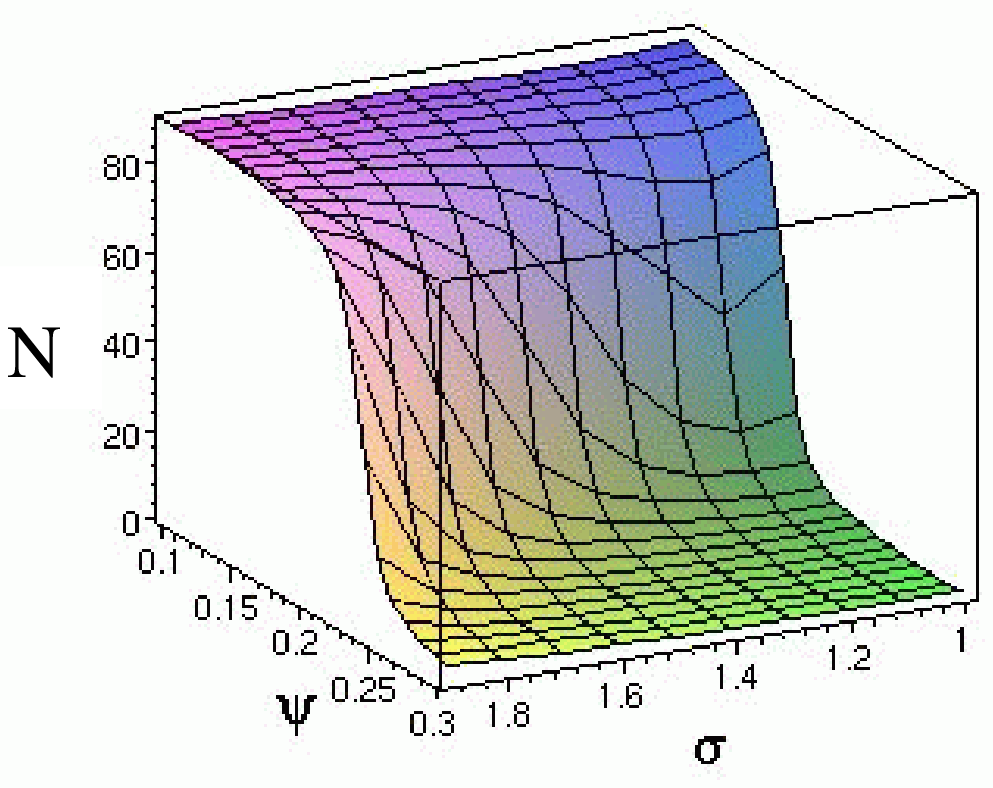,width=1.4\hsize}{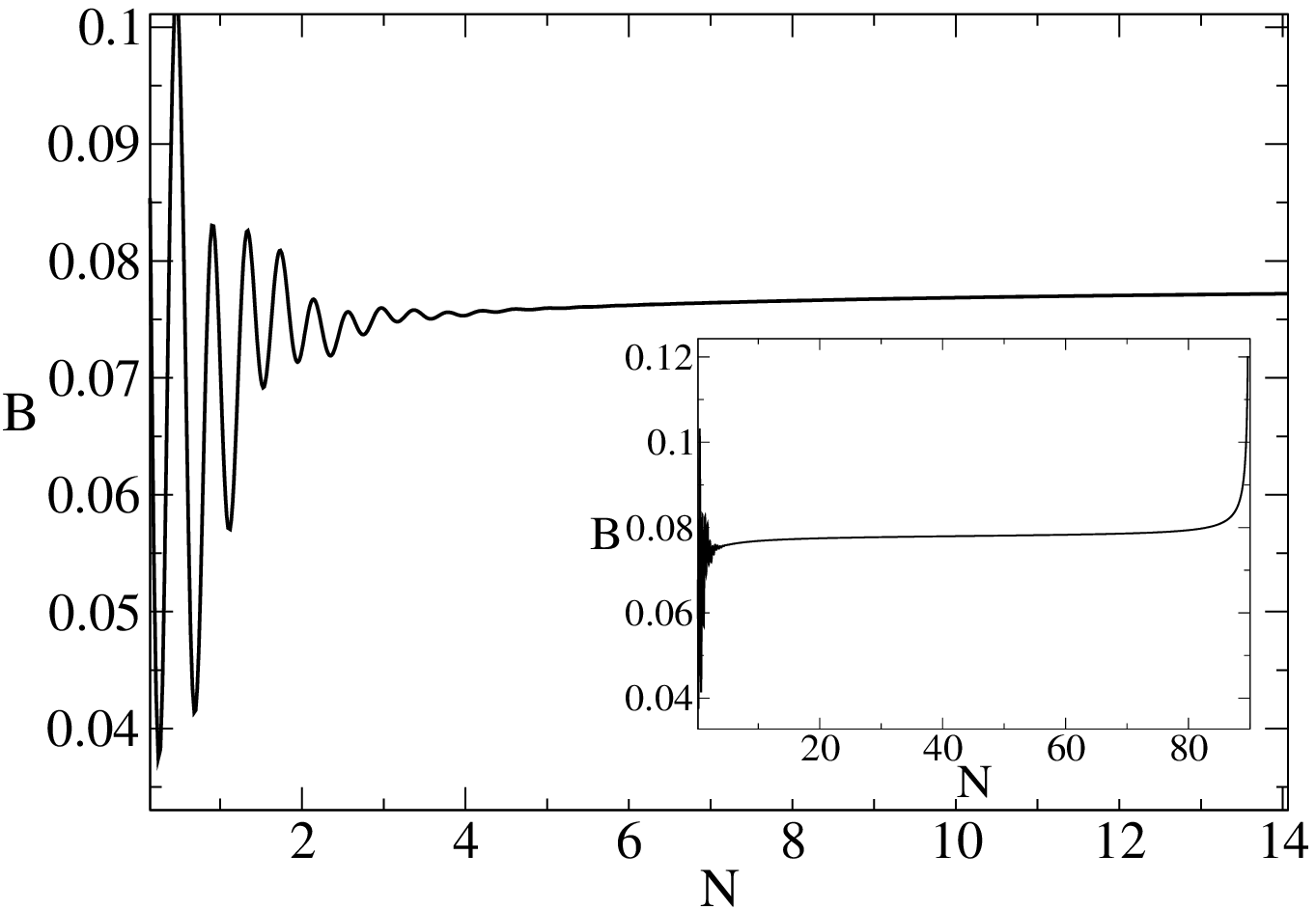,width=1.1\hsize}
{The number of $e$-foldings of expansion as a function of initial
field values for the potential in Fig.\ 6
\label{figefold}}{Evolution of the $B$ field for a typical
trajectory, at early times.  Inset shows the evolution over the
entirety of inflation.\label{figB}}

\medskip\noindent{\it Multi-Field Inflation:}
One might ask whether $\psi$ need always be the inflaton, or
whether instead $\ReT$ could play this role. If so it might be
possible to take advantage of the arguments of ref.~\cite{morebi}
that under certain circumstances a radion such as $\ReT$ can have
a naturally very slow roll. It is also worth searching for
inflationary trajectories where both fields roll, since these can
give rise to observable signals for the Cosmic Microwave
Background (CMB), such as isocurvature density perturbations.

We did not find any examples of radion inflation, but did find
some inflationary trajectories along which both $\ReT$ and
$\ModPhi$ rolled appreciably. For these alternative solutions a
hole can be opened along the side of the trough, allowing the
fields to escape in the direction of increasing $\ReT$, before
finally veering in the direction of brane-antibrane annihilation.
The situation is illustrated in Figs.~\ref{fighole} and
\ref{fighole-closeup}. This behaviour is obtained by lowering the
value of $A$ to 0.897, while keeping other parameters of
(\ref{params}) fixed. Successful inflationary trajectories loiter
at some place in the trough during most of inflation, before
eventually leaving it in the $\ReT$ direction. At the very end
of inflation $\psi$ once again takes over the motion and the
annihilation of brane and antibrane takes place. The twisted
nature of the field-space trajectories is shown in
Fig.~\ref{fignef-curve-hole} (and in a projected view in
Fig.~\ref{figprojected}), which illustrates the evolution of the
fields as a function of the expansion. It is possible to have
periods during which the trajectories curve significantly, which
could have observable consequences for density perturbations, as
we will discuss below.

\DOUBLEFIGURE[ht]{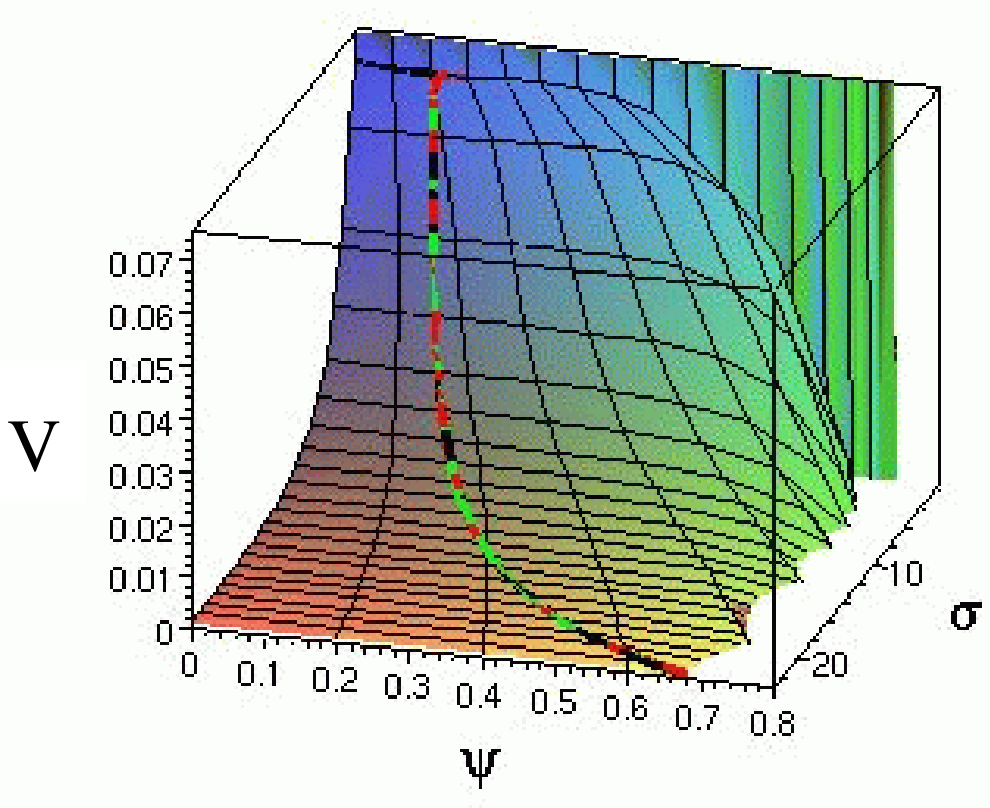,width=1.3\hsize}{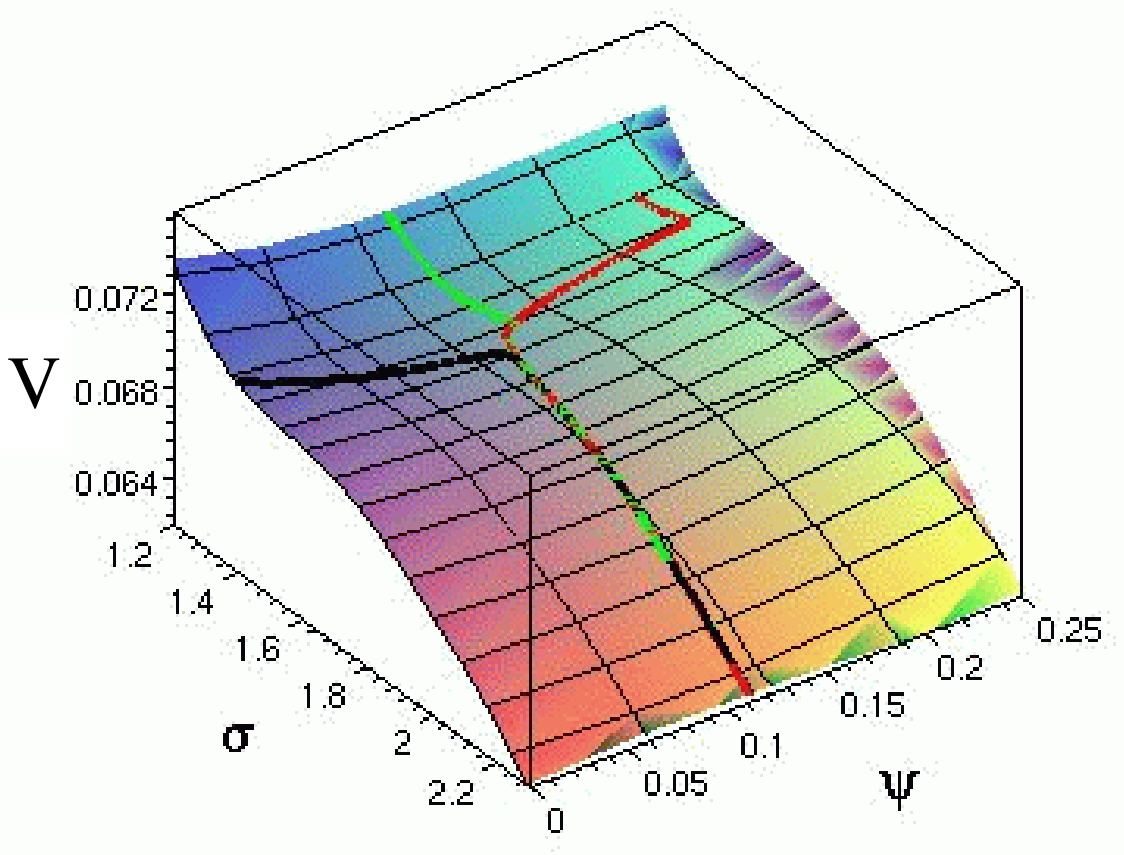,width=1.3\hsize}
{Potential with an opening in the trough, showing several inflaton trajectories.
\label{fighole}}{Closeup of the trough region in figure 10.
\label{fighole-closeup}}

\DOUBLEFIGURE[ht]{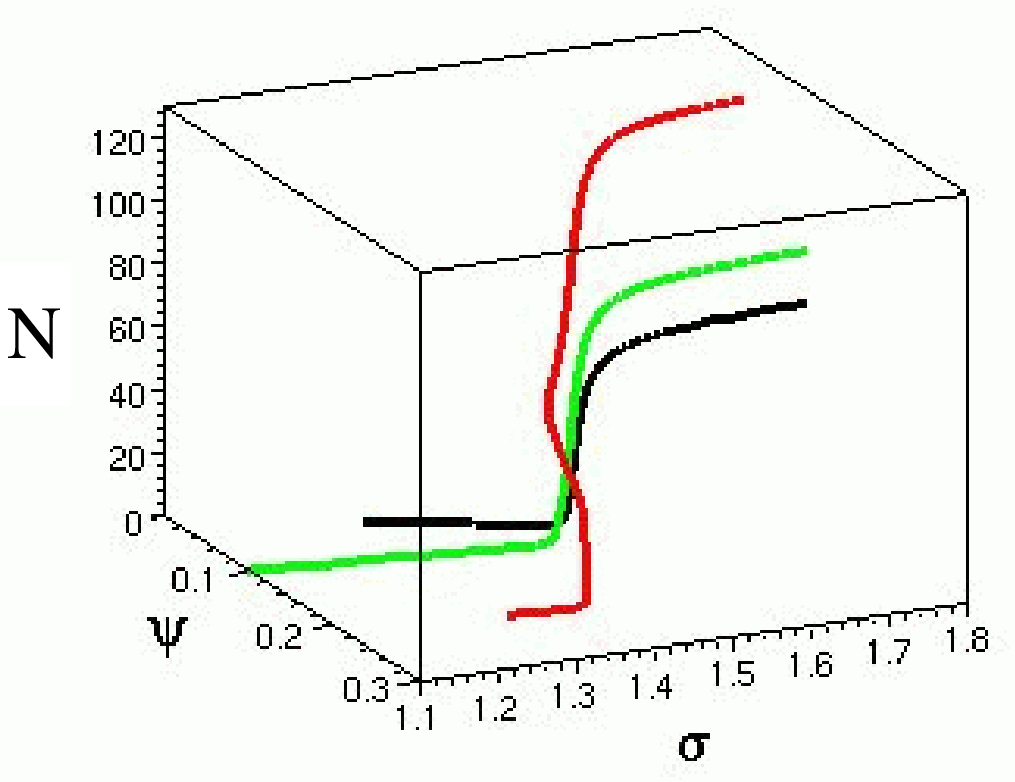,width=1.2\hsize}{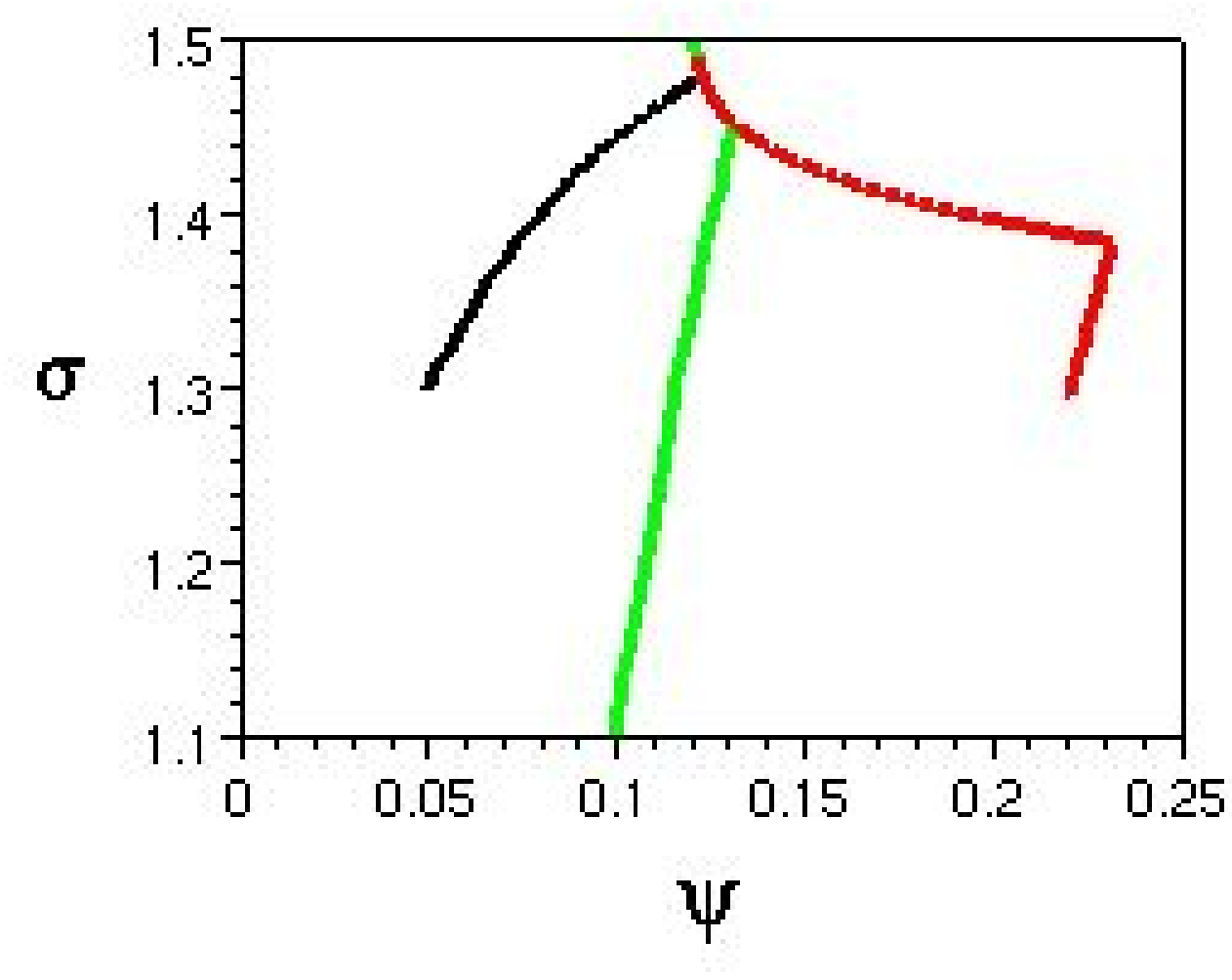,width=1.4\hsize}
{The number of
$e$-foldings versus field values for the three trajectories shown in figures
10 and 11.  \label{fignef-curve-hole}}{Projection of previous figure on
the $\psi$-$\ReT$ plane. \label{figprojected} }

\subsubsection{Scaling Arguments}

The previous examples use a numerically convenient choice of model
parameters which are $O(1)$ in Planck units. We return now to the
issue of whether these solutions lie within the domain of validity
of the low-energy field theory, which require $r, \ReT \gg 1$. We
do so by identifying two separate scaling symmetries which the
solutions to the scalar equations approximately enjoy when they
are in the slow-roll limit.

There is a scale invariance satisfied by {\it any} scalar potential
in the slow-roll approximation, and so which holds throughout
almost the entirety of inflation. This symmetry follows because
the slow-roll equations are unchanged under an overall rescaling
of the potential, $V\to \lambda V$, accompanied by a rescaling of
time, $t\to t/\sqrt{\lambda}$. Under such a change, the slow-roll
equations %%F erased the are
 transform as
\bea  \label{dimensionalscaling}
    H = {\dot a\over a} = \sqrt{V/3}\  &\to&\
     H = \sqrt{\lambda} \, {\dot a\over a} =
    \sqrt{\lambda V/3}; \nonumber\\
    3H\dot\phi^i = -V^{,i}\  &\to& \ 3\sqrt{\lambda}H \sqrt{\lambda}\dot\phi^i =
    -\lambda V^{,i}
\eea
Although the time-dependence of solutions is stretched by this
transformation, the number of $e$-foldings is unchanged and so
there is no change at all in the solutions for the field equations
if it is the number of $e$-foldings, $u \equiv \ln a$, which is
used as the independent variable, $\phi = \phi^i(u)$.

In the present instance this rescaling is accomplished by letting
the Lagrangian parameters scale as,
\be \label{scaling1}
    \{A,C,w\}\to\sqrt{\lambda}\{A,C,w\},\quad \{k,k',1/f_0\}
    \to\lambda\{k,k',1/f_0\} \,,
\ee
which simply corresponds to changing the string scale (in Planck
units). Of course, the freedom to choose this scale is lost once
we demand to reproduce the observed magnitude of the density
perturbations in the CMB, as is done below.

There is a slightly more subtle rescaling property of the
solutions considered here, which has important implications for
the validity of our approximations. In the slow-roll limit, our
scalar equations are unaffected by the transformation:
\be \label{scaling2}
    a\to\lambda' a,\quad \{\ReT,\psi^2,\psi_0^2\}
    \to \{\ReT,\psi^2,\psi_0^2\}
    /\lambda',\quad k' \to k'/\lambda'^2 \,,
\ee
which gives an overall rescaling $V\to \lambda'^2 V$, but without
rescaling the time coordinate. This rescaling to $V$ could be
undone by a transformation of the type (\ref{scaling1}) if
desired, but we instead use it to generate new solutions for which
$\psi$ and $\ReT$ are larger, since $r\to r/\lambda'$. That is, given
any parameter set which leads to a successful inflationary slow
roll, a continuous family of solutions can be constructed that
leads to the same physical predictions, while ensuring that $r$ is
in the range where the effective theory is trustworthy.

\subsection{Density Perturbations}

It is natural to ask for the observable implications of the
inflationary solutions just discussed, so we now calculate the
signature of fluctuations which they predict for the temperature
of the cosmic microwave background (CMB). The precise expression
for the power spectrum in a multi-field inflationary model can be
written as
\be \label{Pexact}
    P(k) = {8\pi G\, V\over 75\pi^2}\, g^{ij}\, {\partial N\over \partial \phi^i}
     {\partial N\over \partial \phi^j}
\ee
(in the notation of %%FF erased this in the notation of
 \cite{Lyth-Riotto}, $P(k) =
\delta_H^2$), where the COBE normalization implies that
$\sqrt{P(k_0)} = 2\times 10^{-5}$ at the scale $k_0 \sim 10^3$
Mpc.

We find that in the present model, the exact power spectrum is
well approximated by the computationally simpler formula
\be
\label{Papprox}
    P(k) = {3\over 75\pi^2}{H^4\over g_{ij}\dot\phi^i\dot\phi^j}
    \,,
\ee
which agrees with (\ref{Pexact}) in the single-field case (where
it also reduces to the familiar expression $\sqrt{P(k)} \sim
H^2/\dot\phi$). The right hand side of these expressions are to be
calculated  at the value of $N$, the number of $e$-foldings since
the beginning of inflation, for which $k=aH=e^N H$.  To the extent
that $H$ is constant during inflation (which is true for our
examples), $P(N)$ has the same functional form as $P(\ln k)$.

In a universe that underwent a total of $N_t$ $e$-foldings of
inflation, only the last 60 or so correspond to fluctuations
within our present horizon.  This number could be lower, depending
on the scale of inflation, which we take to be the string scale
$M_s$, and also on the reheat temperature $T_{\rm rh}$, but if
$T_{\rm rh}\sim M_s\sim 10^{16}$ GeV, then 60 is the expected
number for the $e$-foldings of inflation which have potentially
observable consequences.

The COBE normalization should then be applied at a value near
$N\cong N_t - 60$. Normalizing a typical spectrum obtained from
inflationary trajectories like those shown in Fig.\
\ref{figtrough}, we find that the potential must be rescaled by a
factor of $10^{-11}$ relative to its value corresponding to the
parameters in (\ref{params}). If we assume that these parameters
maintain their order 1 values in units of the string scale rather
than the Planck scale, we then obtain an estimate of the string
scale which would be required to reproduce the observed amplitude
of CMB temperature fluctuations:
\be \label{mstring}
    M_s \cong (10^{-11} )^{1/4} \, M_p \cong 4\times 10^{15}
    \hbox{\ GeV} \, .
\ee

We may similarly ask whether a successful description of the CMB
fluctuations constrains how strongly warped the throat must be.
To the extent that quantities like $k \propto z_b^2$ and $k'
\propto z_b^2 \hat{z}_b^2$ are $O(1)$ in our numerical solutions
during inflation, this also means that the brane tensions are not
strongly suppressed by warping compared to the string scale.

%%CB: These next paragraphs are modified to update the scaling discussion.

When making these estimates we must also return to the issue of
whether the large-$r$ approximation is valid. That is, suppose we
rescale $r$ from its numerically-obtained, $O(1)$, value, $r =
r_0$, by a factor of $\zeta$ to a larger value $r_1 = \zeta \,
r_0$, using transformation \pref{scaling2} with $\lambda' =
1/\zeta$. Then the Lagrangian parameter
$k$ (not to be confused with the wave number of fluctuations!)
remains unchanged by this rescaling but we have
$k' \to \zeta^2 \, k'$ and $V \to V/\zeta^2$. Now, the value of
$V$ can be adjusted back to the phenomenologically successful
value of $10^{-11}$ by raising $V$ by a factor of $\zeta^2$ using
transformation \pref{dimensionalscaling}, with $\lambda =
\zeta^2$. But under such a rescaling both $k$ and $k'$ also
increase by a factor of $\lambda = \zeta^2$, to give $k \to
\zeta^2 \, k$ and $k' \to \zeta^4 \, k'$. If we find that $k$ and
$k'$ obtained after this operation are very small, we may again
conclude the warping is small, but this time within a framework
for which $r$ is acceptably large.

We have searched the parameter space of inflationary solutions,
looking for configurations for which $k$ and $k'$ are small and
the extra-dimensional volume, $r$, is large, in precisely the
above sense. The best values which we found were
\bea
  &&  w= 0.25,\ {f_0}=10^{-2},{A}= 5.5,\  a= 0.45,\ {C}= 1.5\times 10^{-3},
   \nonumber \\ &&  \ c=10^{-2},\ k= 10^{-3},\ k' = 10^{-5}, {\psi_0}= 1.95 
\label{paramsnew} \eea
and for which $\sigma \approx 300$ and $\psi \sim 0.01$ during
inflation. Rescaling this result as above with $\zeta = 1/30$
leaves $r \approx \sigma \approx 10$ throughout inflation, while
rescaling $k \to k/900 \approx 10^{-6}$ and $k' \to k'/810000
\approx 10^{-11}$. The value of $k$ points to a warp factor of
order $z_b \sim 10^{-3}$ at the position of the anti-brane, and
the value $k'/k \approx 10^{-5}$ indicates a warp factor of order
$\hat{z}_b \sim 10^{-5/2} \approx 3 \times 10^{-3}$ at the
position of the mobile D3-brane. This shows that the warping at
the brane and antibrane position is strong enough to
justify our use of approximate formulae based on
both branes being deep within the warped throat. Even so, given a
string scale of $4 \times 10^{15}$ GeV, this implies an antibrane
tension which is about $0.03$ times smaller, $10^{14}$ GeV, and so
which is well above the weak scale.

%%CB: End of changes

The above arguments point to a string scale which is quite close
to the GUT scale because the inflationary roll is not particularly
slow. This has interesting implications for the burning question
of whether the tensor (gravity wave) contribution to the CMB has
any hope of being observed. Current data bound the scale of the
potential to be $V^{1/4} < 3\times 10^{16}$ GeV (see for example
\cite{Cooray}), and it is difficult to push this to lower values
since the figure of merit for observations is $V$, rather than
$V^{1/4}$. Nevertheless, current estimates of the potential for
discovering tensor modes in the CMB indicate that the scale
(\ref{mstring}) is within the reach of future experiments
\cite{Cooray,tensor}.

\DOUBLEFIGURE[ht]{lnPk.eps,width=1.1\hsize}{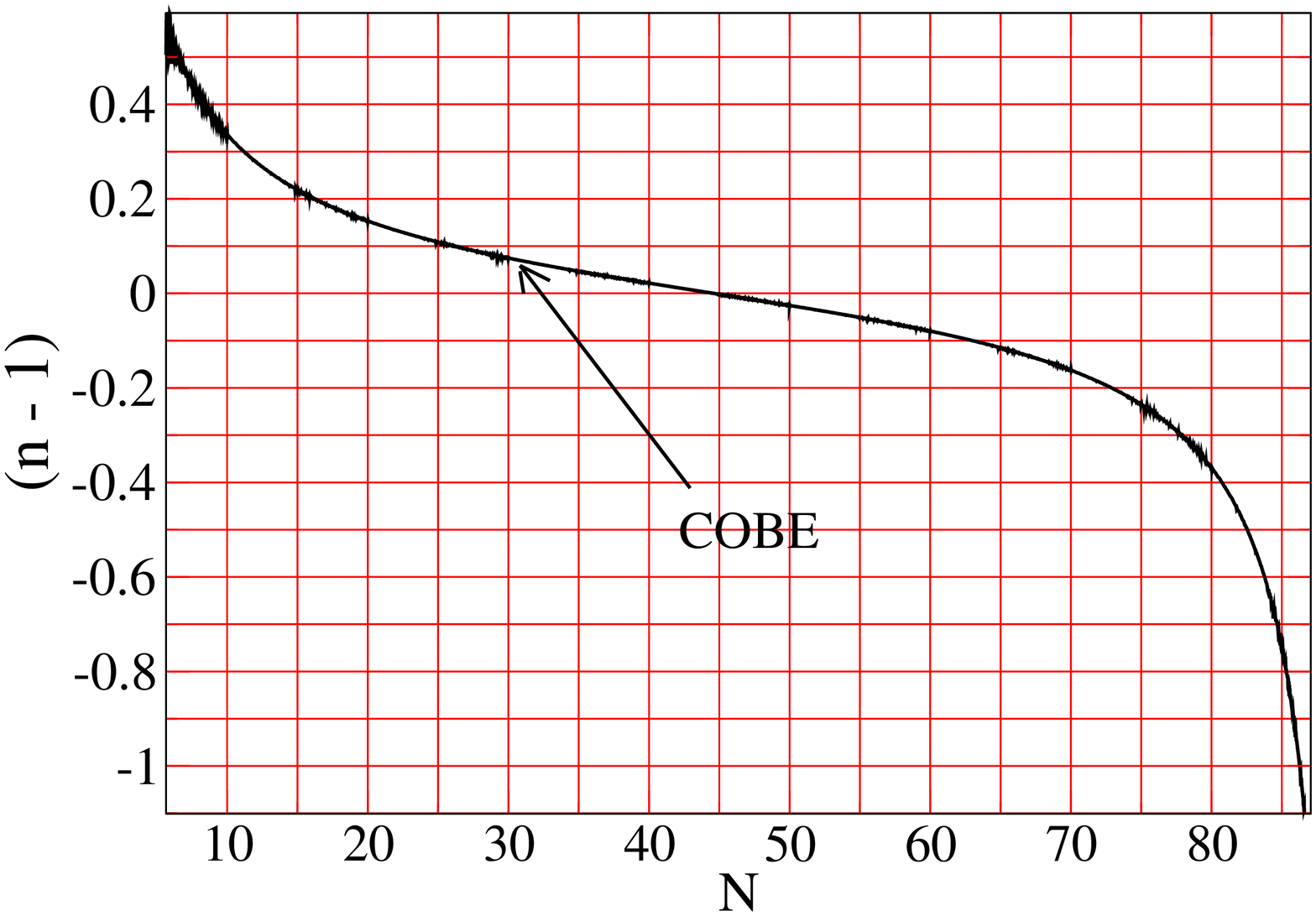,width=1.1\hsize}
{$\ln[P(k)/P_0(k_0)]$ versus number of $e$-foldings since the
beginning of inflation, where $\sqrt{P_0(k_0)} = 2\times 10^{-5}$
is the COBE normalization. \label{figspectrum}}{\label{figindex}
Scalar perturbation index as a function of $N$, related to wave
number through $N\approx\ln k/H$.}

\DOUBLEFIGURE[h]{index-closeup.eps,width=1.1\hsize}{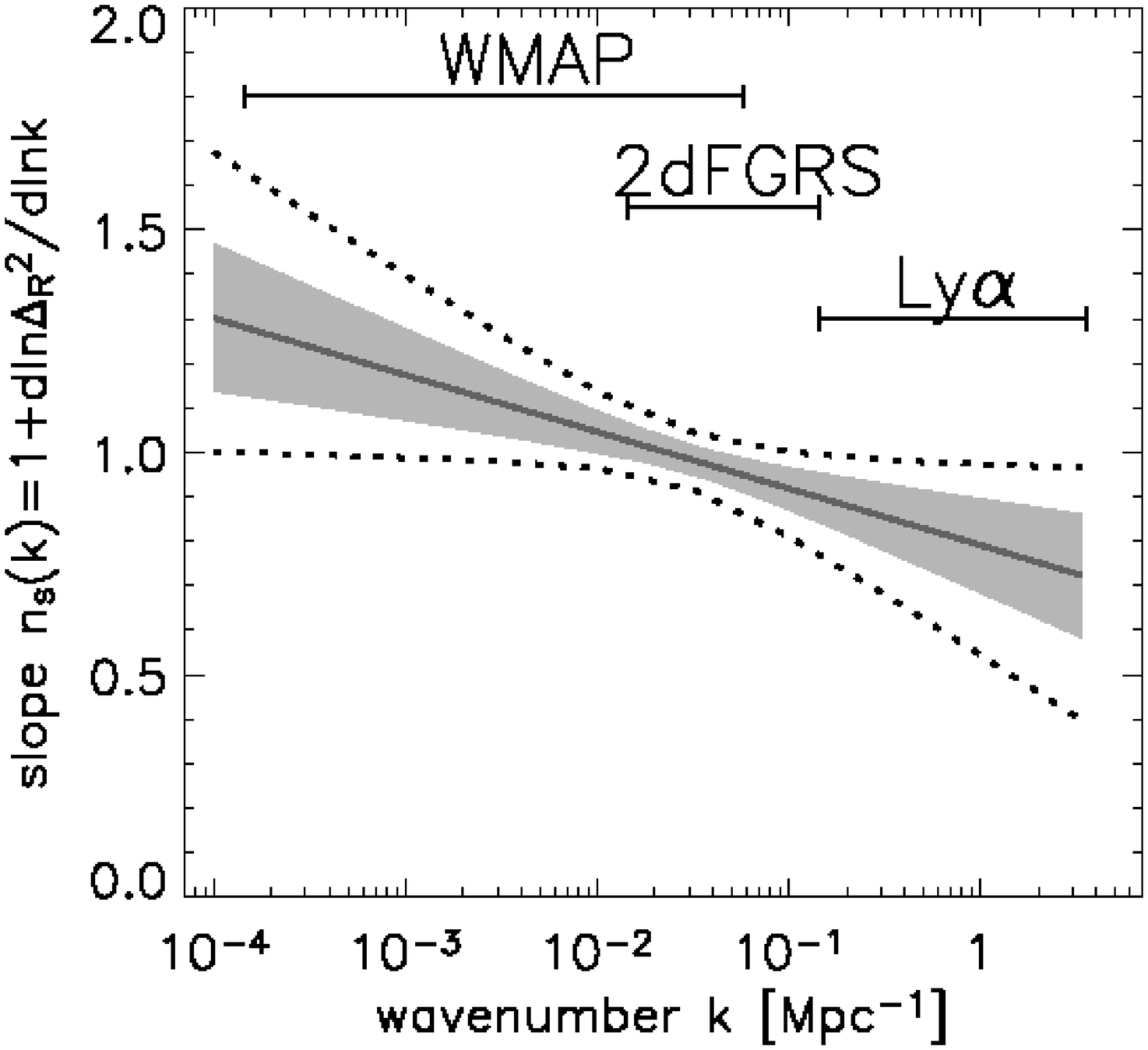,width=1.1\hsize}
{Closeup of the physically interesting region in figure XX, with
$N\sim30-39$ corresponding to wave numbers $k\sim 10^{-4}-1$
Mpc$^{-1}$. \label{figindex-closeup}} {WMAP, LSS and Lyman
$\alpha$ constraints on the spectral index from 68\% (shaded area)
and 95\% (dashed lines) confidence level, from ref.\
\cite{wmap-inf}.\label{figwmap} }

The shape of the spectrum of scalar perturbations is shown in
Fig.\ \ref{figspectrum}. It can be characterized by the spectral
index $n_s$, defined as
\be
    n_s = 1 + {d\ln P(k)\over d\ln k} \,,
\ee
which for our typical solutions is the slope of
Fig.~\ref{figspectrum}.  This is plotted explicitly in
Figs.~\ref{figindex} and \ref{figindex-closeup}. Because it is
difficult to obtain a large amount of inflation, the inflationary
roll is not extremely slow, and the departure from a scale-free
spectrum $(n_s=1)$ tends to be large. In the example shown, the
spectrum is blue in the region relevant for the CMB and large
scale structure formation (shown in Fig.~\ref{figindex-closeup}),
with $n_s\sim 1.03 - 1.08$.

This prediction from brane-antibrane inflation can be compared to
observational constraints from the Wilkinson Microwave Anisotropy
Probe (WMAP), the 2 degree Field Galaxy Redshift Survey (2dFGRS),
and Lyman $\alpha$ forest data, which have been analyzed in ref.\
\cite{wmap-inf}.  Fig.\ \ref{figwmap}, borrowed from fig.\ 2 of
\cite{wmap-inf}, shows that our spectrum is well within the
current limits.  In comparing the prediction with the constraints,
one should identify $N=30$ with $k=10^{-4}$ Mpc$^{-1}$, and $N=39$
with  $k=1$ Mpc$^{-1}$.  Present data are still consistent with a
flat spectrum $n_s=1$ with no running, but there is a suggestion
of large negative running,  $dn/d\ln k \sim -0.04$, with large
error bars.  This hint is reiterated by a recent analysis
combining WMAP data with that of Sloan Digital Sky Survey (SDSS),
which obtained $dn/d\ln k = -0.07\pm 0.04$ \cite{sdss}.  The
probability distribution function is reproduced in Fig.\
\ref{figsdss}.  If the trend toward negative running is confirmed
in future CMB observations at a lower level than the present
central value, it could be a signal in favor of the
brane-antibrane model, which has $dn/d\ln k = -0.01$ in the region
of interest in the example shown. We also plot $dn/d\ln k$ over
the entire inflationary history in  Fig.\ \ref{figdndlnk}. This
shows that larger values of $|dn/d\ln k|$ are indeed correlated
with larger deviations of $n$ from unity, as one would expect from
the theoretical slow-roll expressions for these quantities.

\DOUBLEFIGURE[ht]{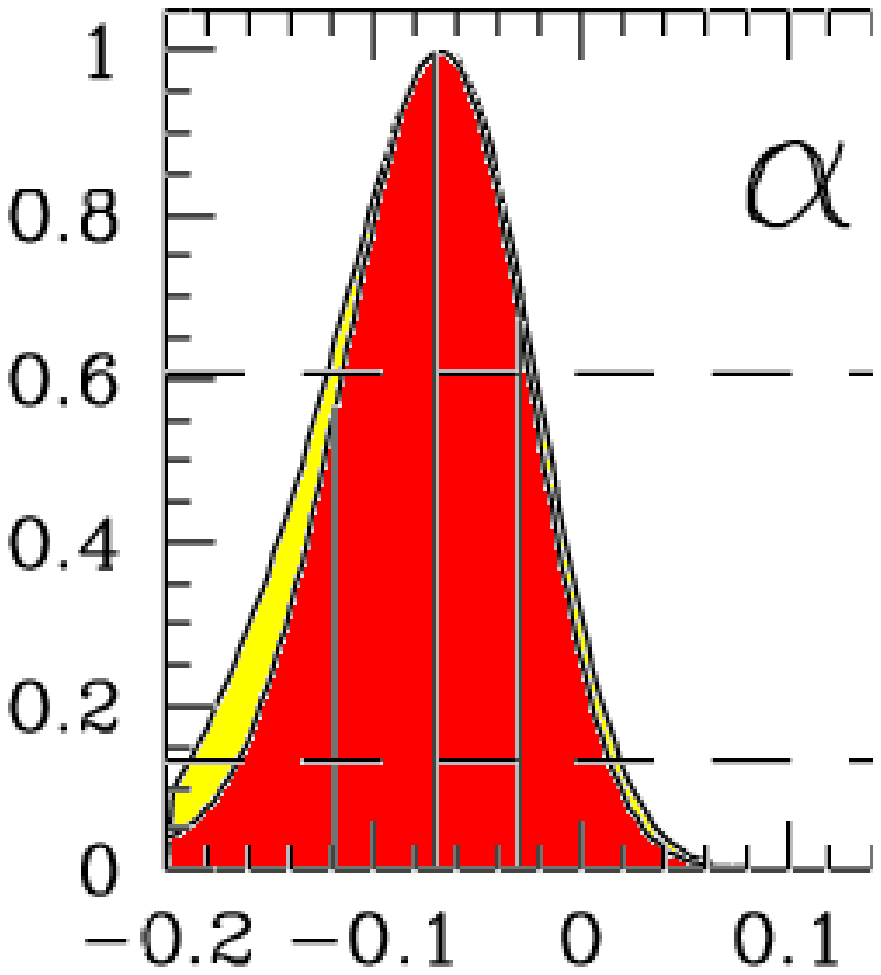,width=0.7\hsize}{dndlnk.eps,width=1.1\hsize}
{Probability distribution function for running of spectral index,
$\alpha\equiv dn/d\ln k$, from combining WMAP and SDSS data \cite{sdss}.
\label{figsdss}}{$dn/d\ln k$ (solid line) and $(n_s-1)/10$ (dashed line) versus $N$
for the inflaton trajectories in the trough.\label{figdndlnk} }

More generally, Fig.\ \ref{figdndlnk} tells us how large a
departure from a pure Harrison-Zeldovich ($n_s=1$) spectrum can be
accommodated in the brane inflation model.  To obtain larger
deviations, the total duration of inflation can easily be
shortened by relaxing the fine tuning of parameters.  The visible
region of the spectrum can thus be moved to lower values of $N$,
leading to deviations as large as $n_s-1=0.5$, $|dn/d\ln
k|=-0.06.$   Although Fig.\ \ref{figdndlnk} corresponds to the
trough trajectories of Fig.\ \ref{figtrough}, where the inflaton
is identified with $\ModPhi$, we have found that the central
$\ReT$-like trajectory of fig.\ \ref{fighole-closeup} produces a
remarkably similar result for both $n_s$ and $dn/d\ln k$.  On the
other hand, interesting variations on this result can be found for
the trajectories neighboring this central one, as shown in Figs.\
\ref{figdndlnk2}-\ref{figdndlnk4}.  The former is less favored by
the data, since it has $dn/d\ln k>0$, but the latter is more
consistent, and provides an example of obtaining distinctive
features in the power spectrum, which could be revealed in future
observations.

\DOUBLEFIGURE[ht]{dndlnk2.eps,width=1.1\hsize}{dndlnk4.eps,width=1.1\hsize}
{$dn/d\ln k$ (solid line) and $(n_s-1)/10$ (dashed line) versus
$N$ for the $\ReT$-like trajectory starting on the right-hand-side
of Fig.\ 11. \label{figdndlnk2}}{Same as previous figure, but
for trajectory starting on the left-hand-side of Fig.\
11.\label{figdndlnk4}}

\medskip\noindent{\it Isocurvature Perturbations:}
In addition to the adiabatic (curvature) perturbations considered
above, the presence of several fields makes it possible to
generate entropy (isocurvature) perturbations--fluctuations in the
light fields  which are orthogonal to the inflaton trajectories.
Isocurvature perturbations alter the shape of the acoustic peaks
of the CMB fluctuation power spectrum, but only if the different
light fields decay after inflation into particles with different
equations of state (such as if $\ModPhi$ decayed into cold dark
matter and $\ReT$ decayed into radiation).  At present there is no
evidence for such perturbations, but only observational bounds on
the level at which they can contribute to the temperature
anisotropy \cite{Julien}.  Here we will not make a detailed
analysis of their potential presence in brane-antibrane inflation;
rather we just point out that some of our inflaton trajectories
fulfill one of the necessary criteria for isocurvature modes to
possibly be observable, namely there must be some curvature in
field space of the inflaton trajectory \cite{GWBM}.  In other
words, the linear combination of fields which constitutes the
inflaton must be time dependent.  This possibility is demonstrated
in Figs.\ \ref{fignef-curve-hole}-\ref{figprojected}, where
significant twisting of the trajectories is evident.  A more
thorough investigation would be warranted if evidence for
contamination by isocurvature modes is found in the data.

\section{Comments and Conclusions}

Our purpose in this paper is to see whether inflation can arise in
the effective theory which captures the essential features of the
low-energy limit of realistic string vacua in which the  moduli
are fixed at the string scale, as in refs.~\cite{gkp,kklt,cmqu}.
Several interesting features emerge from this investigation.

\subsection{`Realistic' Inflation}

Our two main results are these:

\medskip\noindent{\it 1. Explicit Inflationary Solutions:}
We are able to explicitly identify inflationary trajectories
within the 4D effective theory with fixed moduli.
%which captures the low-energy features of string models having most
%of their moduli fixed at the string scale.
In so doing we allow all of the remaining moduli to
roll. We find that inflation appears to be possible in these
models even after moduli stabilization, representing definite
progress over early work \cite{bmnqrz}-\cite{dt} for
which moduli fixing was not addressed. As is usually the case for
inflationary field theories, we find that obtaining a long period of
inflation is not generic
in the sense that it requires tuning of couplings and, to a lesser
extent, initial conditions in field space.

%We find a situation not unlike what is found for inflationary
%field theories: inflation is not generic but is possible for
%special choices of the parameters which describe the effective theory.
In particular, we find that the relevant parameters must be adjusted to 1
part in 1000 in order to obtain 60 $e$-foldings of inflation. Once
these parameters are so chosen, inflation occurs for
a relatively wide range of initial field values (if the various
fields all start from rest). The effective 4D theories have both
$F$-term, $D$-term and supersymmetry-breaking contributions to
their scalar potentials, and we find that all of these terms play
an important role during inflationary evolution.

The inflationary solutions we find could well provide a good
description of the observed CMB temperature fluctuations. Because
it is difficult to obtain a slow roll, the predicted density
fluctuations are typically not deep within the scale invariant
regime. It is quite possible to obtain a scalar index observably
different from unity (on the blue side for the examples
considered), and for which $dn_s/d\ln k$ is different from zero.
Because the inflationary roll is not very slow, observable tensor
perturbations may also be produced.

\medskip\noindent{\it 2. The Standard Model and Reheating:}
We provide the first example of brane-antibrane inflation for
which it is possible to identify the Standard Model degrees of
freedom in the post-inflationary world. This opens up the exciting
possibility of exploring all of the issues associated with
reheating in the post-inflationary universe.

In the models studied in the greatest detail, the Standard Model lives on
an antibrane at the tip of the throat, since this is the choice
we made when using a $U(1)$ gauge coupling which is independent of
$T$. (Presumably we are not living on the antibrane which is
annihilated when the mobile brane reaches the end of the throat.)
Since the tip of the throat can be highly warped, our possible
presence there raises several interesting possibilities. It could
be that the warping along the throat plays a role in the hierarchy
problem, along the lines proposed by Randall and Sundrum. (Of
course, this possibility conflicts with obtaining sufficiently
large perturbations in the CMB within the large $r$ approximation
within the inflationary solutions found here, because these latter
two conditions led us to conclude the total warping should be
small.) %%FF added the next few lines...
It is clearly an open question to obtain a  string
model with a realistic chiral spectrum of quarks and
leptons, with all moduli fixed in such a way that a hierarchy is
naturally obtained after supersymmetry breaking and the scales are
also the ones preferred by the inflation/density perturbation
requirements.
 It would be well worth further exploring the
model-building possibilities along these lines. On the other hand
if the scale preferred by inflation does not match the one needed for
a phenomenologically realistic model after supersymmetry breaking, a
two-throat scenario may be considered in which inflation happens in
one throat with probably not much warping, whereas the standard model
lies on a different throat with enough warping to generate the
hierarchy in the scales.\footnote{We thank Joe Polchinski for
  suggesting this possibility.}

If strong warping could be produced at the throat's tip in a way
consistent with inflation, this might open up other attractive
possibilities for cosmology. In particular, since the warping
tends to reduce the effective tension of the D3 brane as it falls
down the throat, the energy density released by the final
brane-antibrane annihilation is likely to be set by a lower scale
(like the weak scale) rather than the much higher string or Planck
scales. This may be too little to pay the cost of exciting the
comparatively high string-scale masses of states in the bulk or on
branes situated further away from the throat. If so, then the
warping of the throat may act to improve the efficiency with which
inflationary energy gets converted into reheating standard model
degrees of freedom as opposed to populating phenomenologically
problematic bulk states. A full consideration of this process
would be very interesting to pursue but it lies beyond the scope
of the present article. A particularly interesting possibility in
this context is the generation of topological defects
such as cosmic strings {\it after} inflation, such has been
discussed in \cite{cosmicstrings,cmp}. In particular the structure
of the models presented here seems to fit in the class of scenarios
discussed by Copeland, Myers and Polchinski \cite{cmp}, for which
no stable cosmic strings survive.

\subsection{String Theory and Double Inflation}

Although we are able to obtain 60 $e$-foldings of inflation for
some initial conditions, since this is not the generic situation
it is worth standing back and asking whether string theory is
trying to tell us something when it makes inflation not so easy to
achieve.

On reflection there are two things that emerge from the search for
inflation in string theory as being rather generic.

\medskip\noindent$\bullet$
It is difficult to obtain 60 $e$-foldings of inflation at energies
near the string scale, largely because the theory does not have
many small dimensionless numbers with which to work. (In fact the
tunings we required were special values of not particularly small couplings,
rather than unnaturally small values.)  Although
there are many scalar fields which are free to roll at very high
energies, the periods of potential-energy domination which result
are normally not long enough to produce a full 60 $e$-foldings. 10
to 20 $e$-foldings are much easier to obtain however, and perhaps
this suggests that string theory prefers to only give a small
number of $e$-foldings during the inflationary phase which
produces the observed temperature fluctuations in the microwave
background.

\medskip\noindent$\bullet$
String vacua are normally rife with moduli, which generically
acquire masses only after supersymmetry breaks. Thus, there are
likely to be numerous scalar fields whose masses are comparatively small
since they are close to the weak scale, $M_w$. Such scalars
generically cause problems for cosmology since they give rise to a
host of cosmological moduli problems \cite{moduliproblems} during
the Hot Big Bang. Many of these problems would not arise if the
universe were to undergo a period of late-time inflation \cite{lyth-stewart}.

\medskip

Perhaps these two points can lead to a more generic picture of
inflation within string theory. In this picture CMB temperature
fluctuations are produced by an inflationary period involving
energy densities near the string scale, but lasting for only 10 or
more $e$-foldings. The remainder of the 60 $e$-foldings required
to explain the Big Bang's flatness and homogeneity problems arise
during a second period of inflation which is associated with the
rolling of the many string moduli whose masses are of order the
weak scale. For instance the slowness of this later rolling might
be due to a mechanism along the lines proposed in ref.~\cite{dk}.

If this picture is borne out as a {\it bona fide} string prediction,
then it implies several observational consequences.

\begin{itemize}
\item
First, the observed CMB temperature fluctuations should not be
deep into the slow-roll regime because $N_e \ll 60$, and so should
not be extremely close to the scale invariant predictions. In
particular we might expect to find slow-roll parameters $\epsilon$
and $\eta$ which are on the larger side of their allowed ranges,
perhaps being of order $1/N_e \sim 0.1$. Besides more easily
accommodating phenomena like kinks in the inflationary spectrum
and a running spectral index, this would imply that tensor
perturbations might be detected in the near future.
\item
Second, it predicts a period of late inflation, and so requires
any explanation of phenomena like baryon-number generation to
necessarily take place at comparatively low energies like the
electroweak scale.
\end{itemize}

We believe this kind of picture may well represent a more natural
reconciliation between the requirements of inflation and the
properties of known string vacua. If so, it would provide a
natural explanation for effects like a running spectral index,
which may have been observed in the primordial fluctuation
spectrum. We believe more detailed studies of cosmologies of this
sort are warranted given the motivation this kind of picture may
receive both from string theory and the current data.

In summary, we have seen how inflation can arise in an effective
theory which captures the essential features of the low-energy
limit of realistic string vacua with moduli fixed at the string
scale. We believe that we are just seeing the beginnings of the
exploration of inflation in string vacua, and that with the recent
advent of string vacua for which many moduli are fixed at the
string scale \cite{gkp}, much remains to be done towards the goal
of a systematic investigation of the properties of string-based
inflation.

%===================================================================================

\section{Acknowledgements}
We would like to thank J. Blanco-Pilado,  C. Escoda, H. Firouzjahi,
M. G\'omez-Reino,  N. Jones, S. Kachru, R.
Kallosh, A. Linde, J. Maldacena, S. Trivedi,  H. Tye and A. Uranga for helpful
discussions on these and related subjects. We thank the organizers
of the KITP workshop on string cosmology for providing the perfect
environment to start this work. C.B. is funded by NSERC (Canada),
FCAR (Qu\'ebec) and McGill University. F.Q. is partially funded by
PPARC and the Royal Society Wolfson award.

%==================================================================================


\begin{thebibliography}{999}

\bibitem{bmnqrz}
C.~P.~Burgess, M.~Majumdar, D.~Nolte, F.~Quevedo, G.~Rajesh and
R.~J.~Zhang, ``The inflationary brane-antibrane universe,'' JHEP
{\bf 0107} (2001) 047 [hep-th/0105204].

\bibitem{dss}
G.~R.~Dvali, Q.~Shafi and S.~Solganik, ``D-brane inflation,''
[hep-th/0105203].

\bibitem{morebi}
C.~P.~Burgess, P.~Martineau, F.~Quevedo, G.~Rajesh and
R.~J.~Zhang, ``Brane antibrane inflation in orbifold and
orientifold models,'' JHEP {\bf 0203} (2002) 052 [hep-th/0111025].

\bibitem{stephon}
S.~H.~Alexander, ``Inflation from D - anti-D brane annihilation,''
Phys.\ Rev.\ D {\bf 65} (2002) 023507 [hep-th/0105032].

\bibitem{angles}
C.~Herdeiro, S.~Hirano and R.~Kallosh,
``String theory and hybrid inflation / acceleration,''
JHEP {\bf 0112} (2001) 027
[arXiv:hep-th/0110271];
K.~Dasgupta, C.~Herdeiro, S.~Hirano and R.~Kallosh,
``D3/D7 inflationary model and M-theory,''
Phys.\ Rev.\ D {\bf 65} (2002) 126002
[arXiv:hep-th/0203019];
J.~Garcia-Bellido, R.~Rabadan and F.~Zamora, ``Inflationary
scenarios from branes at angles,'' JHEP {\bf 0201} (2002) 036
[hep-th/0112147];
%
N.~Jones, H.~Stoica and S.~H.~Tye, ``Brane interaction as the
origin of inflation,'' JHEP {\bf 0207} (2002) 051
[hep-th/0203163];
%
M.~Gomez-Reino and I.~Zavala, ``Recombination of intersecting
D-branes and cosmological inflation,'' JHEP {\bf 0209} (2002) 020
[hep-th/0207278].

\bibitem{dt}
G.~R.~Dvali and S.~H.~H.~Tye, ``Brane inflation,'' Phys.\ Lett.\ B
{\bf 450} (1999) 72 [hep-ph/9812483].
%%CITATION = HEP-PH 9812483;%%

\bibitem{hybrid}
A.~D.~Linde, ``Hybrid inflation,'' Phys.\ Rev.\ D {\bf 49} (1994)
748 [astro-ph/9307002].

\bibitem{sen}
For a recent discussion including many references, see:
A.~Fotopoulos and A.~A.~Tseytlin, ``On open superstring partition
function in inhomogeneous rolling tachyon background,''
hep-th/0310253;
A.~Sen,
``Remarks on tachyon driven cosmology,''
arXiv:hep-th/0312153.

\bibitem{kklmmt}
S.~Kachru, R.~Kallosh, A.~Linde, J.~Maldacena, L.~McAllister and
S.~P.~Trivedi, ``Towards inflation in string theory,''
hep-th/0308055.

\bibitem{jellium}
S.~Buchan, B.~Shlaer, H.~Stoica and S.~H.~H.~Tye, ``Inter-brane
interactions in compact spaces and brane inflation,''
hep-th/0311207.

\bibitem{kklt}
S. Kachru, R. Kallosh, A. Linde and S. P. Trivedi, ``de Sitter
Vacua in String Theory,'' [hep-th/0301240].

\bibitem{etaprob}
See for instance: E.~J.~Copeland, A.~R.~Liddle, D.~H.~Lyth,
E.~D.~Stewart and D.~Wands, ``False vacuum inflation with Einstein
gravity,'' Phys.\ Rev.\ D {\bf 49} (1994) 6410 [astro-ph/9401011].
%%CITATION = ASTRO-PH 9401011;%%

\bibitem{dterm}
P.~Binetruy and G.~R.~Dvali, ``D-term inflation,'' Phys.\ Lett.\ B
{\bf 388} (1996) 241 [hep-ph/9606342].

\bibitem{KS} 
T.~Kobayashi and O.~Seto,
``Dilaton and moduli fields in D-term inflation,''
Phys.\ Rev.\ D {\bf 69}, 023510 (2004)
[arXiv:hep-ph/0307332];
%%CITATION = HEP-PH 0307332;%%
T.~Higaki, T.~Kobayashi and O.~Seto,
``D-term inflation and nonperturbative Kaehler potential of dilaton,''
arXiv:hep-ph/0402200.
%%CITATION = HEP-PH 0402200;%%

\bibitem{cmqu}
J.~F.~G.~Cascales, M.~P.~Garcia del Moral, F.~Quevedo and
A.~M.~Uranga, ``Realistic D-brane models on warped throats:
Fluxes, hierarchies and moduli stabilization,'' hep-th/0312051.

\bibitem{rs}
L.~Randall and R.~Sundrum, ``A large mass hierarchy from a small
extra dimension,'' Phys.\ Rev.\ Lett.\  {\bf 83} (1999) 3370
[hep-ph/9905221].

\bibitem{verlinde}
H.~Verlinde, ``Holography and compactification,'' Nucl.\ Phys.\ B
{\bf 580} (2000) 264 [hep-th/9906182].
%%CITATION = HEP-TH 9906182;



\bibitem{sethi}
S.~Sethi, C.~Vafa and E.~Witten,
``Constraints on low-dimensional string compactifications,''
Nucl.\ Phys.\ B {\bf 480} (1996) 213
[arXiv:hep-th/9606122];
K.~Dasgupta, G.~Rajesh and S.~Sethi,
``M theory, orientifolds and G-flux,''
JHEP {\bf 9908} (1999) 023
[arXiv:hep-th/9908088].

\bibitem{gkp}
S.~B.~Giddings, S.~Kachru and J.~Polchinski, ``Hierarchies from
fluxes in string compactifications,'' Phys. Rev. {\bf D66}, 106006
(2002).

\bibitem{gvw}
S. Gukov, C. Vafa and E. Witten, ``CFTs from Calabi-Yau
Fourfolds,'' Nucl. Phys. {\bf B584}, 69 (2000).

\bibitem{truncation}
E.~Witten, ``Dimensional Reduction Of Superstring Models,'' Phys.\
Lett.\ B {\bf 155} (1985) 151;
%
C.~P.~Burgess, A.~Font and F.~Quevedo, ``Low-Energy Effective
Action For The Superstring,'' Nucl.\ Phys.\ B {\bf 272} (1986)
661.

\bibitem{noscale} E. Cremmer, S. Ferrara, C. Kounnas and
D.V. Nanonpoulos, ``Naturally vanishing cosmological constant in
$N=1$ supergravity,'' Phys. Lett. {\bf B133}, 61 (1983);
%
J. Ellis, A.B. Lahanas, D.V. Nanopoulos and K. Tamvakis,
``No-scale Supersymmetric Standard Model,'' Phys. Lett. {\bf
B134}, 429 (1984).

\bibitem{gauginocondensation}
J.~P.~Derendinger, L.~E.~Ibanez and H.~P.~Nilles, ``On The
Low-Energy D = 4, N=1 Supergravity Theory Extracted From The D =
10, N=1 Superstring,'' Phys.\ Lett.\ B {\bf 155} (1985) 65;
%%CITATION = PHLTA,B155,65;%%
%
M.~Dine, R.~Rohm, N.~Seiberg and E.~Witten, ``Gluino Condensation
In Superstring Models,'' Phys.\ Lett.\ B {\bf 156} (1985) 55.
%%CITATION = PHLTA,B156,55;%%

\bibitem{ourgc}
C.P. Burgess, J.-P. Derendinger, F. Quevedo and M. Quir\'os,
``Gaugino Condensates and Chiral-Linear Duality: An
Effective-Lagrangian Analysis'', Phys.\ Lett.\ B {\bf 348} (1995)
428--442;
%%CITATION = PHLTA,B348,428;%%
``On Gaugino Condensation with Field-Dependent Gauge Couplings'',
Ann.\ Phys.\ {\bf 250} (1996) 193-233.

\bibitem{egq}
C.~Escoda, M.~Gomez-Reino and F.~Quevedo, ``Saltatory de Sitter
string vacua,''JHEP {\bf 0311} (2003) 065, hep-th/0307160.

\bibitem{bkq}
C.~P.~Burgess, R.~Kallosh and F.~Quevedo, ``de Sitter string vacua
from supersymmetric D-terms,'' JHEP {\bf 0310} (2003) 056
[hep-th/0309187].

\bibitem{eva}
A. Saltman and E. Silverstein, ``The Scaling of the No Scale
Potential and de Sitter Model Building,'' [hep-th/0402135].

\bibitem{Brustein:2004xn}
R.~Brustein and S.~P.~de Alwis,
 ``Moduli potentials in string compactifications with fluxes:
 Mapping the discretuum,''
arXiv:hep-th/0402088.

\bibitem{nonlinsusy}
J.~Hughes and J.~Polchinski, ``Partially Broken Global
Supersymmetry And The Superstring,'' Nucl.\ Phys.\ B {\bf 278}
(1986) 147.
%
R.~Altendorfer and J.~Bagger, ``Dual supersymmetry algebras from
partial supersymmetry breaking,'' Phys.\ Lett.\ B {\bf 460} (1999)
127 [hep-th/9904213];
%
C.~P.~Burgess, E.~Filotas, M.~Klein and F.~Quevedo, ``Low-energy
brane-world effective actions and partial
 supersymmetry breaking,'' JHEP {\bf 0310} (2003) 041 [hep-th/0209190].

\bibitem{Giddings}
O. De Wolfe and S.B. Giddings, ``Scales and Hierarchies in Warped
Compactifications and Brane Worlds,'' Phys.\ Rev.\ {\bf D67}
(2003) 066008 [hep-th/0208123].

\bibitem{BL}
C.P. Burgess and C.A. L\"utken, Phys.\ Lett.\ {\bf B153} (1985)
137.

\bibitem{hkp}
J.~P.~Hsu, R.~Kallosh and S.~Prokushkin, ``On brane inflation with
volume stabilization,'' JCAP {\bf 0312} (2003) 009
[hep-th/0311077];
\bibitem{Kallosh:2004rs}
R.~Kallosh and S.~Prokushkin,
``SuperCosmology,''
arXiv:hep-th/0403060.
%%CITATION = HEP-TH 0403060;%%

\bibitem{henry}
H.~Firouzjahi and S.~H.~H.~Tye, ``Closer towards inflation in
string theory,'' hep-th/0312020.
%%CITATION = HEP-TH 0312020;%%

\bibitem{Buchel:2003qj}
A.~Buchel and R.~Roiban,
``Inflation in warped geometries,''
arXiv:hep-th/0311154;
%%CITATION = HEP-TH 0311154;%%
E.~Halyo,
``D-brane inflation on conifolds,''
arXiv:hep-th/0402155.

\bibitem{riotto}
L.~Pilo, A.~Riotto and A.~Zaffaroni, ``Old inflation in string
theory,'' hep-th/0401004.

\bibitem{smbranes}
G.~Aldazabal, L.~E.~Ibanez, F.~Quevedo and A.~M.~Uranga,
``D-branes at singularities: A bottom-up approach to the string
embedding of the standard model,'' JHEP {\bf 0008} (2000) 002
[hep-th/0005067].

\bibitem{dsw}
M.~Dine, N.~Seiberg and E.~Witten, ``Fayet-Iliopoulos Terms In
String Theory,'' Nucl.\ Phys.\ B {\bf 289} (1987) 589.

\bibitem{bdkv}
P. Binetruy, G. Dvali, R. Kallosh and A. Van Proeyen,
[hep-th/0402046].

\bibitem{fi}
P.~Fayet and J.~Iliopoulos, ``Spontaneously Broken Supergauge
Symmetries And Goldstone Spinors,'' Phys.\ Lett.\ B {\bf 51}
(1974) 461.
%%CITATION = PHLTA,B51,461;%%

\bibitem{NLvsExpl}
J.~M.~Cornwall, D.~N.~Levin and G.~Tiktopoulos, ``Derivation Of
Gauge Invariance From High-Energy Unitarity Bounds On The S -
Matrix,'' Phys.\ Rev.\ D {\bf 10} (1974) 1145 [Erratum-ibid.\ D
{\bf 11} (1975) 972];
%
C.~P.~Burgess and D.~London, ``Uses and abuses of effective
Lagrangians,'' Phys.\ Rev.\ D {\bf 48} (1993) 4337
[hep-ph/9203216].
%%CITATION = HEP-PH 9203216;%%

\bibitem{ciu}
P.~G.~Camara, L.~E.~Ibanez and A.~M.~Uranga, ``Flux-induced
SUSY-breaking soft terms,'' [hep-th/0311241].
%%CITATION = HEP-TH 0311241;%%

\bibitem{Doug}
See for instance: S.~Groot Nibbelink and B.~J.~W.~van Tent,
``Density perturbations arising from multiple field slow-roll
inflation,'' [hep-ph/0011325];
%%CITATION = HEP-PH 0011325;%%
%
C.~P.~Burgess, P.~Grenier and D.~Hoover, ``Quintessentially flat
scalar potentials,'' [hep-ph/0308252].
%%CITATION = HEP-PH 0308252;%%

\bibitem{CFM}
J.~M.~Cline, H.~Firouzjahi and P.~Martineau, ``Reheating from
tachyon condensation,'' JHEP {\bf 0211}, 041 (2002)
[hep-th/0207156];
%%CITATION = HEP-TH 0207156;%%
J.~M.~Cline and H.~Firouzjahi,
``Real-time D-brane condensation,''
Phys.\ Lett.\ B {\bf 564}, 255 (2003)
[arXiv:hep-th/0301101];
%%CITATION = HEP-TH 0301101;%%
N.\ Barnaby and J.~M.~Cline, in preparation

\bibitem{Lyth-Riotto}
D.~H.~Lyth and A.~Riotto, ``Particle physics models of inflation
and the cosmological density perturbation,'' Phys.\ Rept.\  {\bf
314}, 1 (1999) [hep-ph/9807278].
%%CITATION = HEP-PH 9807278;%%

\bibitem{Cooray}
A.~Cooray, ``After MAP: Next generation CMB,'' [astro-ph/0211347].
%%CITATION = ASTRO-PH 0211347;%%

\bibitem{tensor}
L.~Knox and Y.~S.~Song, ``A limit on the detectability of the
energy scale of inflation,'' Phys.\ Rev.\ Lett.\  {\bf 89}, 011303
(2002) [astro-ph/0202286];
%%CITATION = ASTRO-PH 0202286;%%
%
M.~Kesden, A.~Cooray and M.~Kamionkowski, ``Separation of
gravitational-wave and cosmic-shear contributions to  cosmic
microwave background polarization,'' Phys.\ Rev.\ Lett.\  {\bf
89}, 011304 (2002) [astro-ph/0202434].
%%CITATION = ASTRO-PH 0202434;%%
%
W.~H.~Kinney, ``The energy scale of inflation: Is the hunt for the
primordial B-mode a waste of time?,'' [astro-ph/0307005].
%%CITATION = ASTRO-PH 0307005;%%

\bibitem{wmap-inf}
H.~V.~Peiris {\it et al.}, ``First year Wilkinson Microwave
Anisotropy Probe (WMAP) observations: Implications for
inflation,'' Astrophys.\ J.\ Suppl.\  {\bf 148}, 213 (2003)
[astro-ph/0302225].
%%CITATION = ASTRO-PH 0302225;%%

\bibitem{sdss}
M.~Tegmark {\it et al.}  [SDSS Collaboration], ``Cosmological
parameters from SDSS and WMAP,'' [astro-ph/0310723].
%%CITATION = ASTRO-PH 0310723;%%

\bibitem{Julien}
P.~Crotty, J.~Garcia-Bellido, J.~Lesgourgues and A.~Riazuelo,
``Bounds on isocurvature perturbations from CMB and LSS data,''
Phys.\ Rev.\ Lett.\  {\bf 91}, 171301 (2003) [astro-ph/0306286];
M.~Bucher, J.~Dunkley, P.~G.~Ferreira, K.~Moodley and C.~Skordis,
``The initial conditions of the universe: how much isocurvature is allowed?,''
arXiv:astro-ph/0401417.
%%CITATION = ASTRO-PH 0401417;%%
%%CITATION = ASTRO-PH 0306286;%%

\bibitem{GWBM}
C.~Gordon, D.~Wands, B.~A.~Bassett and R.~Maartens, ``Adiabatic
and entropy perturbations from inflation,'' Phys.\ Rev.\ D {\bf
63}, 023506 (2001) [astro-ph/0009131].
%%CITATION = ASTRO-PH 0009131;%%

\bibitem{kpv}
S. Kachru, J. Pearson and H. Verlinde, ``Brane/Flux Annihilation
and the String Dual of a Non-Supersymmetric Field Theory,'' JHEP
{\bf 0206}, 021 (2002) [hep-th/0112197].

\bibitem{krasnikov}
N.~V.~Krasnikov, ``On Supersymmetry Breaking In Superstring
Theories,'' Phys.\ Lett.\ B {\bf 193} (1987) 37.

\bibitem{moduliproblems}
G.~D.~Coughlan, W.~Fischler, E.~W.~Kolb, S.~Raby and G.~G.~Ross,
``Cosmological Problems For The Polonyi Potential,'' Phys.\ Lett.\
B {\bf 131} (1983) 59;
%
B.~de Carlos, J.~A.~Casas, F.~Quevedo and E.~Roulet, ``Model
independent properties and cosmological implications of the
dilaton and moduli sectors of 4-d strings,'' Phys.\ Lett.\ B {\bf
318} (1993) 447 [hep-ph/9308325];
%%CITATION = HEP-PH 9308325;%%
%
T.~Banks, D.~B.~Kaplan and A.~E.~Nelson, ``Cosmological
implications of dynamical supersymmetry breaking,'' Phys.\ Rev.\ D
{\bf 49} (1994) 779 [hep-ph/9308292].
%%CITATION = HEP-PH 9308292;%%


\bibitem{cosmicstrings}
N.~T.~Jones, H.~Stoica and S.~H.~H.~Tye, ``The production,
spectrum and evolution of cosmic strings in brane inflation,''
Phys.\ Lett.\ B {\bf 563} (2003) 6 [hep-th/0303269];
%
L.~Pogosian, S.~H.~H.~Tye, I.~Wasserman and M.~Wyman,
``Observational constraints on cosmic string production during
brane inflation,'' Phys.\ Rev.\ D {\bf 68} (2003) 023506
[hep-th/0304188];
%
M.~Majumdar and A.~Christine-Davis,
``Cosmological creation of D-branes and anti-D-branes,''
JHEP {\bf 0203} (2002) 056
[arXiv:hep-th/0202148];
%
G.~Dvali and A.~Vilenkin, ``Formation and evolution of cosmic
D-strings,'' hep-th/0312007; G.~Dvali, R.~Kallosh and A.~Van
Proeyen, ``D-term strings,'' JHEP {\bf 0401} (2004) 035
[hep-th/0312005].

\bibitem{cmp}
E.~J.~Copeland, R.~C.~Myers and J.~Polchinski, ``Cosmic F- and
D-strings,'' hep-th/0312067.

\bibitem{lyth-stewart}
D.~H.~Lyth and E.~D.~Stewart,
``Thermal inflation and the moduli problem,''
Phys.\ Rev.\ D {\bf 53}, 1784 (1996)
[arXiv:hep-ph/9510204].
%%CITATION = HEP-PH 9510204;%%

\bibitem{dk}
G.~Dvali and S.~Kachru, ``New old inflation,'' hep-th/0309095.
%%CITATION = HEP-TH 0309095;%%


\end{thebibliography}
\end{document}